\documentclass[aps,prc,letterpaper,11pt,twoside,tightenlines,nofootinbib,showpacs,preprint]{revtex4}
\usepackage{graphicx}
\usepackage[sort&compress]{natbib}
\begin{document}

\newcommand{\eg}{{\it e.g.}}
\newcommand{\etal}{{\it et. al.}}
\newcommand{\ie}{{\it i.e.}}
\newcommand{\be}{\begin{equation}}
\newcommand{\dd}{\displaystyle}
\newcommand{\ee}{\end{equation}}
\newcommand{\bea}{\begin{eqnarray}}
\newcommand{\eea}{\end{eqnarray}}
\newcommand{\bef}{\begin{figure}}
\newcommand{\eef}{\end{figure}}
\newcommand{\bce}{\begin{center}}
\newcommand{\ece}{\end{center}}
\def\lsim{\mathrel{\rlap{\lower4pt\hbox{\hskip1pt$\sim$}}
    \raise1pt\hbox{$<$}}}         
\def\gsim{\mathrel{\rlap{\lower4pt\hbox{\hskip1pt$\sim$}}
    \raise1pt\hbox{$>$}}}         

\title{Hadronic Modes and Quark Properties in the Quark-Gluon Plasma}

\author{M.~Mannarelli}
\altaffiliation{Present address: Center for Theoretical Physics,
Laboratory for Nuclear Science and Department of Physics
\\Massachusetts Institute of Technology, Cambridge, MA 02139.}
\affiliation{Cyclotron Institute and Physics Department, Texas A\&M
University, College Station, Texas 77843-3366}
\author{R.~Rapp}
\affiliation{Cyclotron Institute and Physics Department, Texas A\&M
University, College Station, Texas 77843-3366}

\date{\today}

\begin{abstract}
Based on interaction potentials between a heavy quark and antiquark
as extracted from recent QCD lattice calculations, we set
up a Brueckner-type many-body scheme to study the properties of
light (anti-) quarks in a Quark-Gluon Plasma at moderate
temperatures, $T\simeq$~1-2~$T_c$. The quark-antiquark $T$-matrix,
including both color-singlet and -octet channels, and corresponding
quark self-energies and spectral functions are calculated
self-consistently. The repulsive octet potential induces
quasiparticle masses of up to 150~MeV, whereas the attractive
color-singlet part generates resonance structures in the $q$-$\bar
q$ $T$-matrix, which in turn lead to quasiparticle widths
of $\sim$200~MeV. This corresponds to scattering rates
of $\sim$1~fm$^{-1}$ and may reflect liquid-like properties of the
system.
\end{abstract}
\preprint{MIT-CTP 3639}
  \maketitle

\section{Introduction}
A central goal of the relativistic heavy-ion collision program is
the creation and identification of new forms of highly excited
nuclear matter, in particular a deconfined and chirally symmetric
Quark-Gluon Plasma (QGP). At sufficiently high temperature $T$, due
to asymptotic freedom of Quantum Chromodynamics (QCD), the QGP is
expected to be a weakly interacting gas of quark- and
gluon-quasiparticles with comparatively small thermal masses,
$m_{q,g}\sim gT$. Recent data from the Relativistic Heavy-Ion
Collider (RHIC) indicate, however, that the produced matter exhibits
strong collective behavior which is incompatible with a weakly
interacting QGP: standard (2~$\leftrightarrow$~2) perturbative QCD
(pQCD) cross sections for quarks and gluons do not allow for rapid
thermalization~\cite{Baier:2000sb} as
required in hydrodynamic models to reproduce the observed magnitude
of the elliptic flow~\cite{Kolb:2003dz,Teaney:2000cw,Hirano:2004er}.
With estimated initial energy densities well in excess of the
critical one predicted by lattice QCD (lQCD),
$\epsilon_c\simeq1~$GeV/fm$^3$, the question arises what the nature 
of the produced medium at temperatures $T\simeq$~1-2~$T_c$ is
($T_c\simeq$~170~MeV: critical temperature). Of particular
importance is the identification of the relevant interactions that
can lead to sufficiently large scattering rates while maintaining
consistency with the QGP equation of state (EoS), as determined in
lQCD.

Recent (quenched) lQCD calculations found intriguing evidence that
mesonic correlation functions, after transformation into Minkowski
space, exhibit resonance (or bound-state) like structures for
temperatures up to $\sim$$2\,T_c$. This was first observed for
low-lying charmonia ($\eta_c$,
$J/\psi$)~\cite{Datta:2002ck,Asakawa:2003re,Umeda:2002vr}, but
subsequently also for mesonic systems with lighter
quarks~\cite{Karsch:2003jg,Asakawa:2003nw}. As is well known,
resonance scattering is typically characterized by isotropic angular
distributions and thus more efficient in randomizing momentum
distributions than forward-dominated pQCD cross sections. Indeed, a
recent calculation~\cite{vanHees:2004gq} based on the assumption of
resonant ``$D$"-meson states in the QGP has shown that thermal
relaxation times for charm quarks are reduced by a factor of $\sim$3
as compared to using perturbative rescattering cross sections. The
possibility of light hadronic states (especially for the pion and its
chiral partner $\sigma$) surviving above the phase transition has been
suggested some time ago using effective quark interactions, \eg, within
the Nambu-Jona-Lasinio model~\cite{Hatsuda:1984jm,Hatsuda:1985eb},
within the instanton-liquid model based on euclidean
correlators~\cite{Schafer:1995df}, or more recently in
Refs.~\cite{Li:2004ue,Alberico:2004we}.

To make closer contact to lQCD, some recent works have extracted
a (color-singlet) heavy-quark ($Q$-$\bar Q$) potential, $V_1$, from
the corresponding lQCD free energy, $F_1$, at finite $T$, and injected
it into a Schr\"odinger equation to infer quarkonium
properties~\cite{Digal:2001ue,Wong:2004zr,Mocsy:2004bv}. Reasonable
consistency was found in that the heavy-quark bound states
dissolve at roughly the same temperatures at which the peaks in the
lQCD spectral functions disappear ($\sim$$2\,T_c$ for $J/\psi$ and
$\eta_c$), provided the free energy was converted into a potential
by subtracting an entropy term according to $V_1=F_1 - T~dF_1/dT$.
A similar approach has also been applied to the light-quark sector
in Refs.~\cite{Shuryak:2003ty,Brown:2003km,Shuryak:2004tx}, where the 
$q$-$\bar q$ potentials from unquenched lQCD (including colored 
channels) have been supplemented by relativistic (and instanton-induced) 
interaction corrections. Assuming rather large quark- and 
gluon-quasiparticle masses, $m_{q,g}\simeq$~3-4~$T$ (motivated by lQCD 
calculations of temporal masses~\cite{Petreczky:2001yp}), 
light mesonic, as well as a large number
of colored diquark, quark-gluon and gluon-gluon, bound states have been
found. Both quark-/gluon-quasiparticles and binary bound states 
together were shown to approximately reproduce the
EoS from lQCD. However, the effects of finite widths for both
(anti-) quarks and bound states, which are essential to address
scattering problems, were not included.

In the present article we employ quark-antiquark potentials
extracted from lQCD (including relativistic corrections as in
Refs.~\cite{Shuryak:2003ty,Brown:2003km,Shuryak:2004tx})
within a 3-dimensionally reduced Bethe-Salpeter equation to
evaluate (anti-) quark interactions in the QGP. We compute the
pertinent scattering ($T$-) matrices in both color-singlet
and -octet channels and calculate the quark self-energies including
both real and imaginary parts (corresponding to quasiparticle masses
and widths). The self-energies, in turn, are reinserted into the
$q$-$\bar q$ propagator of the $T$-matrix equation, constituting a
self-consistency problem which we solve by numerical iteration. We
comment on possible consequences of our results for quasiparticle
masses and widths with respect to the QGP EoS  and (anti-) quark
rescattering timescales, respectively.

Our article is organized as follows. In Sec.~\ref{sec_latt} we
present our parametrization of lQCD data for the singlet free energy
and extract a pertinent quark-antiquark potential including both
color-singlet and -octet contributions. In Sec.~\ref{sec_BS} we set
up our self-consistency problem comprising the $q$-$\bar q$ scattering
equation and in-medium single particle self-energies and propagators,
and discuss the underlying assumptions and approximations. The
numerical results with accompanying discussion for the $T$-matrix
and self-energy in a nonperturbative QGP are contained in
Sec.~\ref{sec_res}. In Sec.~\ref{sec_concl} we conclude and give an
outlook.

\section{Quark-Antiquark Potential From Lattice QCD}
\label{sec_latt}
To obtain a driving term (potential) for a $q$-$\bar q$ scattering
equation we take recourse to lQCD calculations of the static free
energy for a $Q$-$\bar Q$ pair. The Bielefeld group has performed
extensive studies of this quantity based on Polyakov loop
correlators~\cite{Philipsen:2002az} for both the pure-glue
$SU(3)$~\cite{Kaczmarek:2002mc,Kaczmarek:2004gv} and
$N_f$=2-QCD~\cite{Kaczmarek:2003dp,Petreczky:2004priv}. Various
parameterizations thereof have been given in the literature, cf.,
\eg, Refs.~\cite{Karsch:1987pv,Petreczky:2004pz,Wong:2004zr,Digal:2004}.
\begin{figure}[!th]
\includegraphics[height=3.in,width=3.0in,angle=-90]{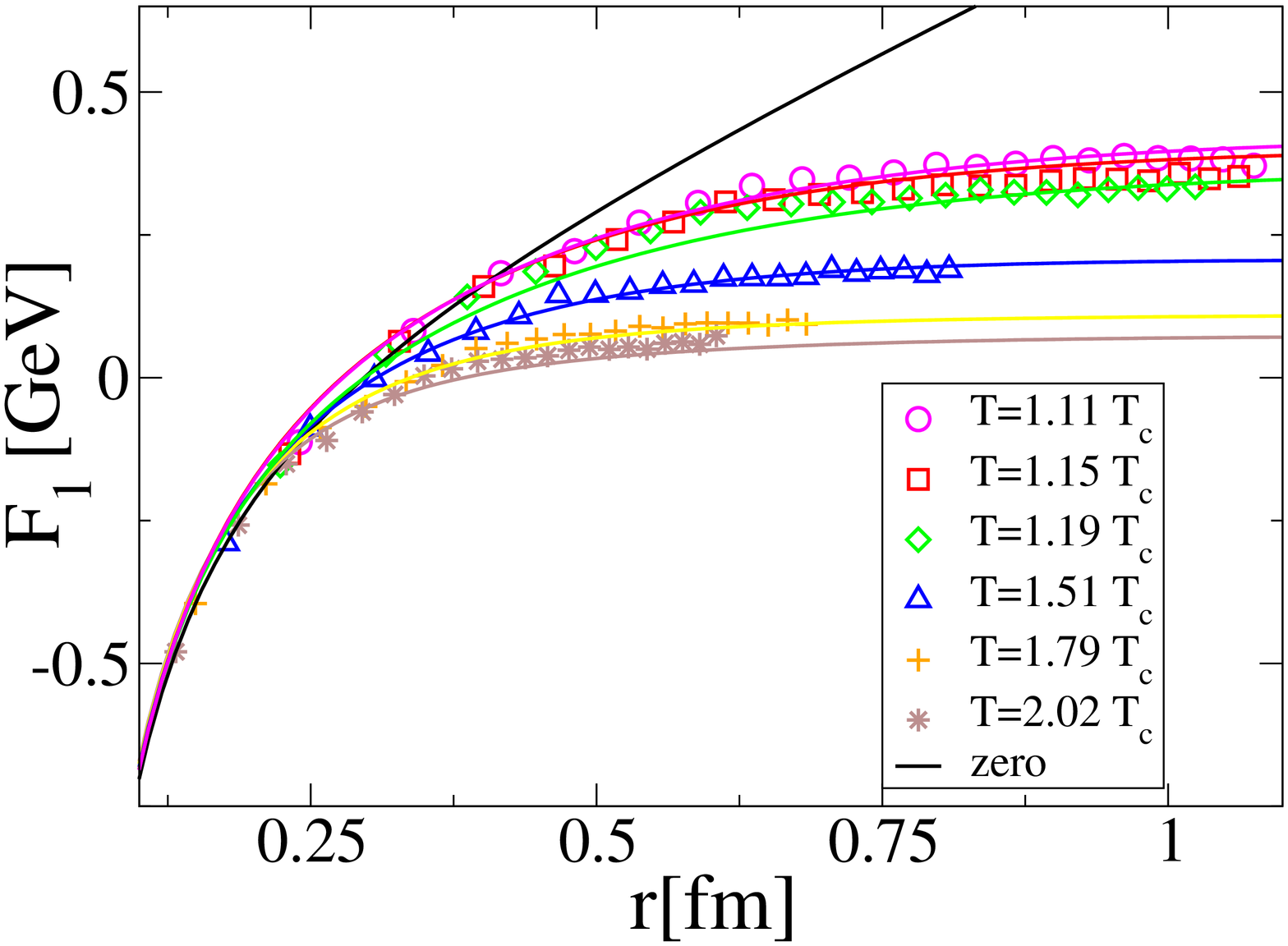}
\includegraphics[height=3.in,width=3.0in,angle=-90]{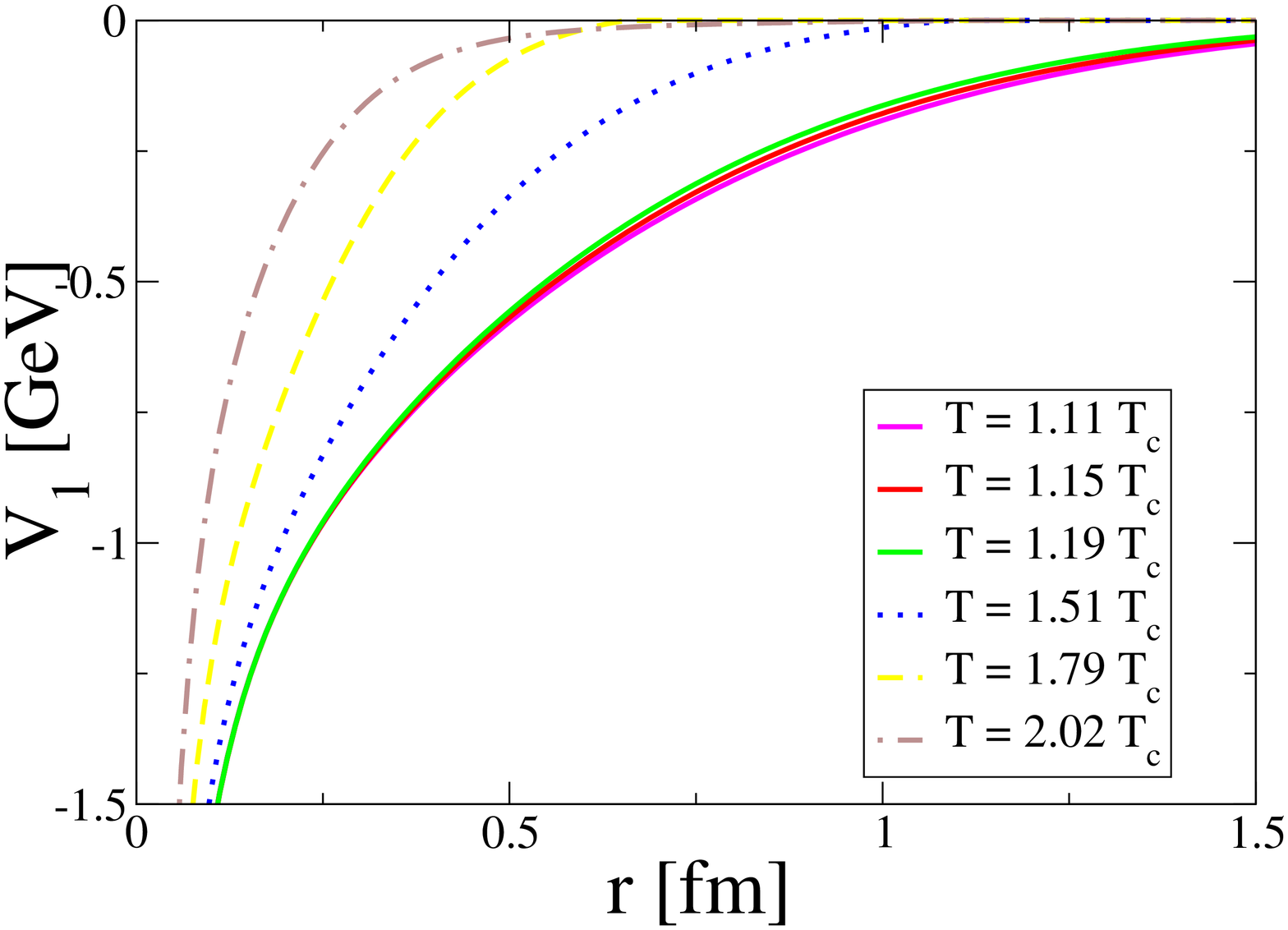}
\caption{Left panel: lattice QCD results for the color-singlet free
energy from unquenched simulations~\cite{Petreczky:2004priv}  for 6
different values of the temperature (symbols) compared to our fit
function, Eq.~(\ref{singlet}), represented by the various curves.
Right panel: corresponding potential in the color-singlet channel
obtained with Eq.~(\ref{V_1}) for the 6 different values of the
temperature.} \label{fig_lat}
\end{figure}
For the temperature range $T=$~1.1-2~$T_c$, it turns out that unquenched
singlet free energy~\cite{Kaczmarek:2003dp,Petreczky:2004priv}
can be reasonably well reproduced by the the following
form reminiscent of a screened Cornell potential (as suggested,
\eg, in Ref.~\cite{Karsch:1987pv}),
\be
F_1(r,T) = -\frac{\alpha}{r}e^{-a \mu(r,T) r} +
        \frac{\sigma}{\mu(r,T)}(1-e^{-\mu(r,T) r }) \, ,
\label{singlet} \ee with a ``screening mass" 
\be \mu(r,T) =
\frac{\sigma}{b} e^{-0.3/r} \,  \label{screening} 
\ee 
and two fitting functions given  by 
\be a \equiv a(r,T) =
\frac{1}{2\sqrt{\mu(r,T)}} \hspace*{1. cm} b \equiv b(t)= 1.1
-3.6\,t -4.3\,t^2 + 17.5\,t^3 
\ee 
where  $t= T/T_c$, $\alpha=0.4$ and $\sigma=1.2 \,{\rm GeV}^2$. 
The left panel of Fig.~\ref{fig_lat}
summarizes our fit to the lattice ``data". Also shown is the
unquenched zero-temperature potential as obtained in
Ref.~\cite{Petreczky:2004pz} (recall that for $T$=0, $E_1=F_1$, see
also below), which is used to normalize the finite-$T$ results at
short distances, $r < $~0.2~fm, where the free energy is not
expected to depend on temperature anymore. Our parametrization,
Eq.~(\ref{singlet}), accommodates this $T$=0 constraint.

As mentioned in the Introduction, the appropriate quantity in
relation to the free
energy that can serve as an effective potential appears to be the
(color-singlet) internal energy $E_1$. Following Kaczmarek
\etal~\cite{Kaczmarek:2004gv}, we subtract the entropy contribution
to the free energy according to
\be E_1\,=\,F_1-T\frac{d F_1}{d T} \,.
\ee
The nonzero asymptotic value of the internal energy can now
be interpreted as an in-medium quark mass that should not be
included in the interaction part of the potential. One therefore
assumes that the  potential in the color-singlet channel can be
extracted via
\be V_1(r,T)=E_1(r,T) -E_1(\infty,T) \, . \label{V_1}
\ee
The singlet potential is shown in the right panel of Fig.~\ref{fig_lat}
for the same values of temperature as the singlet
free-energy (left panel).  The potentials are appreciably larger in
magnitude than the corresponding free energies and decrease with
increasing temperature.

\begin{figure}[!b]
\includegraphics[height=3.in,width=3.0in,angle=-90]{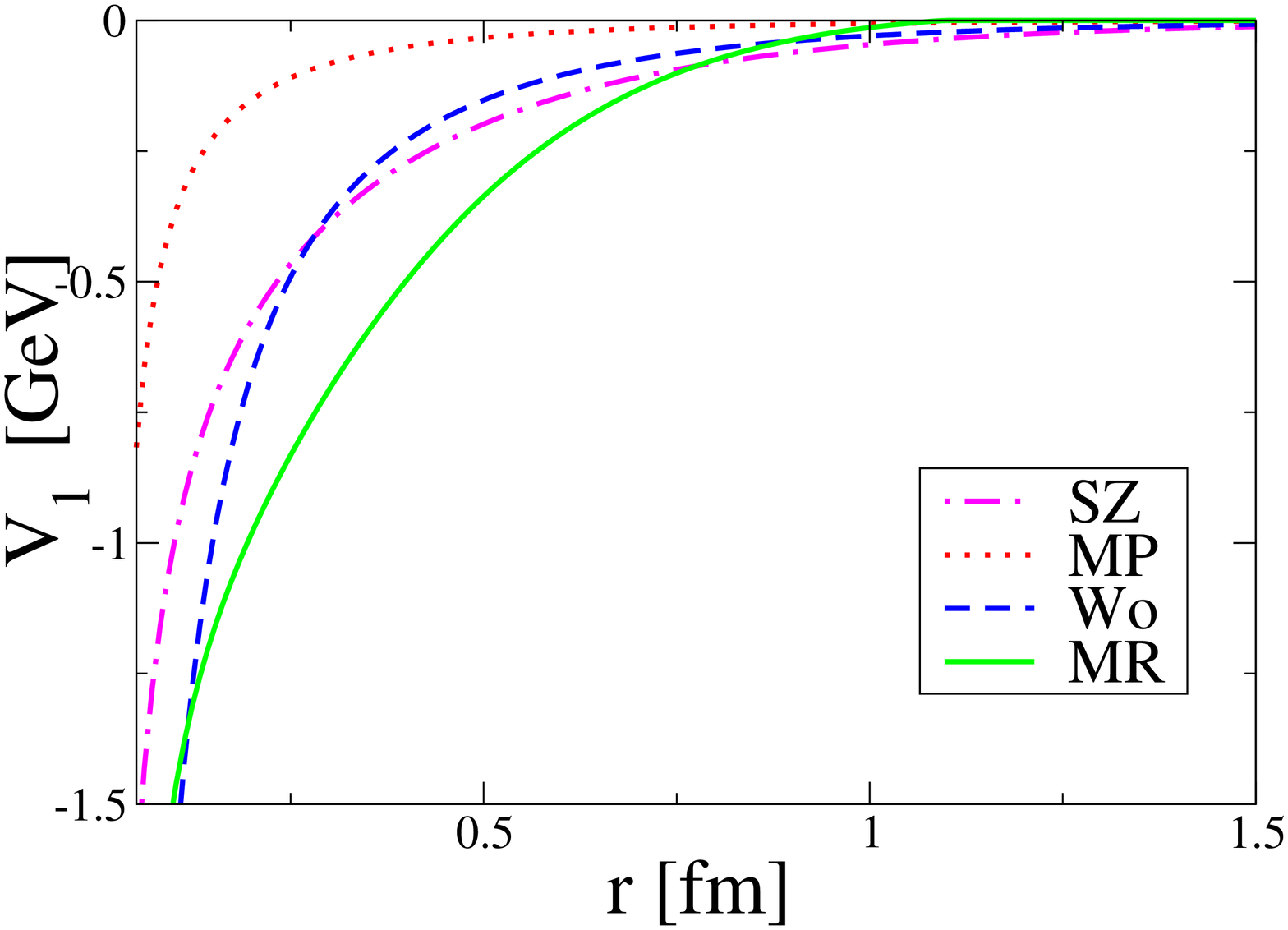}
\includegraphics[height=3.in,width=3.0in,angle=-90]{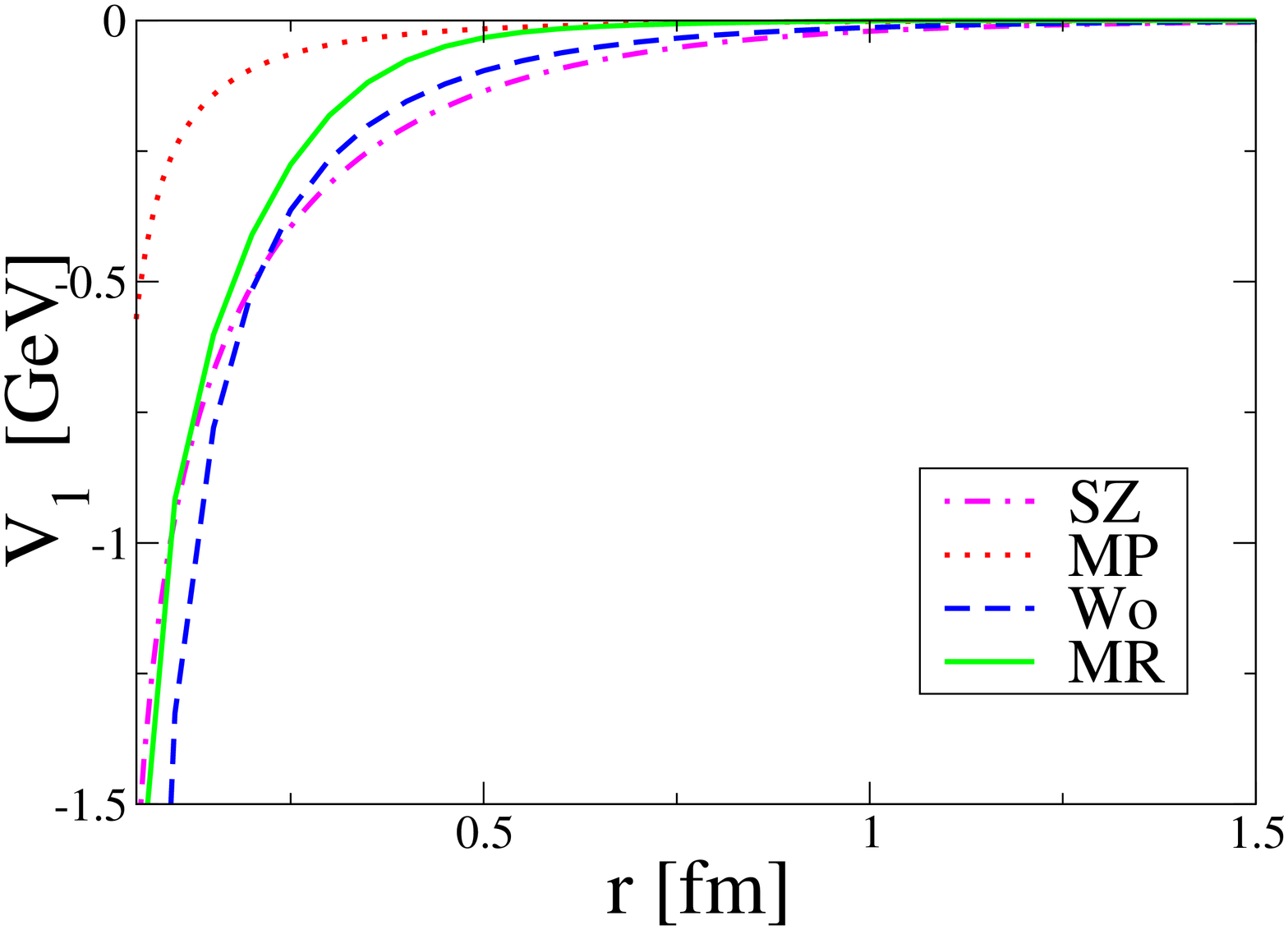}
\caption{Comparison of $Q$-$\bar Q$ color-singlet potentials 
as determined from various lattice data in our (MR) and other 
works (SZ~\cite{Shuryak:2004tx},
MP~\cite{Mocsy:2004bv} and Wo~\cite{Wong:2004zr}); the left (right)  
panel is for $T=1.5\,T_c$~$(2\,T_c)$.} 
\label{fig_comp}
\end{figure}

To illustrate uncertainties in the determination of the potentials
we compare in Fig.~\ref{fig_comp} our results with the ones obtained 
by other groups for temperatures of 1.5~$T_c$ (left panel)
and 2~$T_c$ (right panel). While the potentials of 
Refs.~\cite{Wong:2004zr} (Wo) and \cite{Mocsy:2004bv} (MP) are 
extracted from quenched lQCD, the one
of Ref.~\cite{Shuryak:2004tx} (SZ) and ours (MR) result from
unquenched simulations. 
At 1.5~$T_c$ our potential is about 30-40\% more attractive than SZ at 
distances between 0.1-0.8~fm, while
similar deviations also occur between the quenched-based results,
while at 2~$T_c$ our potential is less attractive than both
the quenched (Wo) and unquenched (SZ) results.
We therefore conclude that the current uncertainty in the extraction
of the potentials amounts to about 50\%, and that it does not yet
allow for a systematic discrimination between quenched and unquenched 
results (provided the temperature dependence is normalized to $T_c$, 
which is, of course, quite different in quenched ($\sim$260~MeV) and 
unquenched ($\sim$170~MeV) simulations). The pertinent uncertainties
will be assessed below by performing $T$-matrix calculations
for charmonium ($c$-$\bar c$ systems) with both our and the quenched 
Wo potential (cf.~Sect.~\ref{subTmatrix}), and comparing them to
spectral functions from lQCD~\cite{Asakawa:2003re} obtained with
different methods; reasonable agreement will be found.

Toward a more complete description of the $q$-$\bar q$ interactions
in the QGP we will in this work also consider the (repulsive)
contributions from the color-octet channel. However, as pointed out
in Ref.~\cite{Jahn:2004qr}, the octet potential cannot be
straightforwardly inferred from the Polyakov loop correlators. Due
to a lack of better knowledge of the octet free energies, we here
assume that the octet potential follows the leading-order result of
perturbation theory,
\be F_8 = - \frac{1}{8} F_1 \, . \label{V8}
\ee
Again, we will check the sensitivity of our calculations to this
approximation, by varying the coefficient in Eq.~(\ref{V8}) by a
factor of 0.5-2. 

For non-static quarks it is also important to include relativistic
corrections~\cite{Brown:1952ph,Brown:2003km}. Following
Ref.~\cite{Brown:2003km} we implement a velocity-velocity
interaction term by the replacement $ V(r) \to V(r)(1 - \hat\alpha_1
\cdot \hat\alpha_2 )$ where $\hat\alpha_1$ and $\hat\alpha_2$ are
quasiparticle velocity operators. As pointed out in
Ref.~\cite{Shuryak:2004tx}, this procedure is strictly speaking
correct only for a Coulomb-type potential.

\section{Reduced Bethe-Salpeter Equation, Quark Self-Energy and
Self-Consistency} \label{sec_BS}
To evaluate quark-antiquark interactions in the QGP we employ the
$T$-matrix approach, as is well known from the nuclear many-body
problem. In relativistic field theory, the starting point is a
4-dimensional Bethe-Salpeter (BS) equation,
\be
 T = K + \int K S S T \, ,
\label{BS0}
\ee
where $K$ denotes the interaction kernel and $S$ is
the single-particle propagator. Both quantities carry, in principle,
dependencies on temperature and (baryon-) density of the surrounding
medium. Since the effective $q$-$\bar q$ potential
constructed in the previous section is essentially non-relativistic
in nature, it is appropriate to employ the ladder
approximation to Eq.~(\ref{BS0}) in connection with neglecting
virtual particle-antiparticle loops. Thus, we will identify the
kernel $K$ with the potential $V$ with appropriate approximations in
the propagator and scattering equation to be discussed in the
following.

\begin{figure}[t]
\includegraphics[height=3.in,width=5.0in,angle=0]{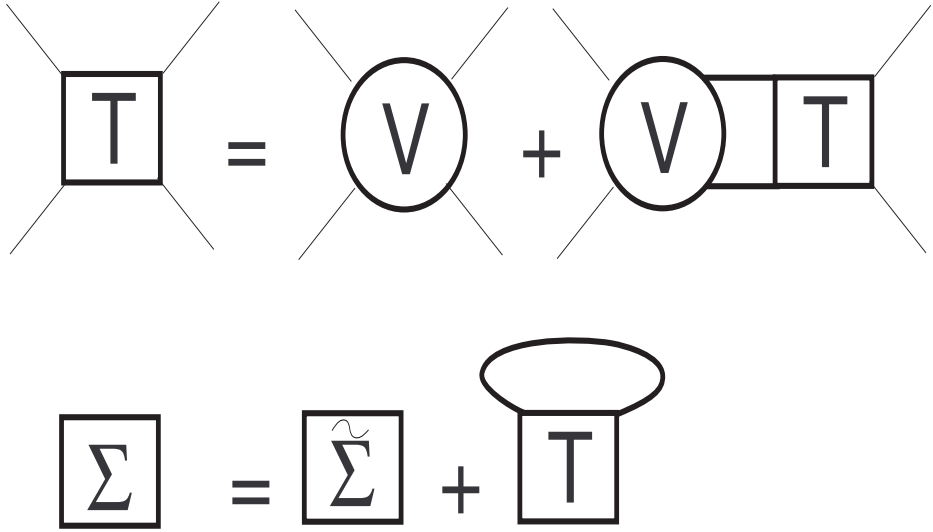}
\caption{Schematic representation of the self-consistency
problem composed of the Bethe-Salpeter equation (\ref{BS0}) in ladder
approximation and the quark self-energy, Eq.~(\ref{Sigma0}).
Thick lines represent full fermionic propagators.
The four different blocks correspond to the $T$-matrix ($T$),
potential ($V$), self-energy ($\Sigma$) and ``gluon-induced"
self-energy ($\tilde\Sigma$).}
\label{fig_BSSD}
\end{figure}

The medium effects in the quark propagator, $S$, are encoded in a
self-energy which we decompose according to
\be \Sigma = \tilde \Sigma
+\int\! T S
 \label{Sigma0} \, .
\ee
The first term, $\tilde\Sigma$, represents a ``gluon-induced"
contribution due to interactions of (anti-) quarks with
surrounding thermal gluons. In this work we do not calculate
this term explicitly, but we will study
how different (purely real) values affect our results.
We note that a perturbative
(hard-thermal-loop) form of this (mass-) term is widely used as
a parameter in quasiparticle descriptions of the QGP
EoS~\cite{Levai:1997yx,Schneider:2001nf,Peshier:2002ww,Blaizot:2003tw}.
The second term on the right side of Eq.~(\ref{Sigma0}) is the
contribution to the self-energy induced by interactions with
antiquarks of the heat bath which we compute at the same
level of approximation as the $T$-matrix.
In principle, the quark self-energy also receives contributions from
interactions with thermal quarks (which could be significant especially in
the scalar diquark channel), but we neglect them in this work.
We also constrain ourselves to the case of vanishing quark chemical
potential, $\mu_q=0$, which implies equal self-energies for quarks
and antiquarks. The two equations (\ref{BS0}) and (\ref{Sigma0})
constitute a self-consistency problem which is diagrammatically
illustrated in Fig.~\ref{fig_BSSD}.

Let us discuss the single-particle quantities in more detail.
The full quark propagator obeys a Schwinger-Dyson equation,
\be
 S =  S_0 + S_0 \Sigma S \,,
\label{SchwingerDyson} \ee where $S_0(k)=[k\!\!\!/-m_0]^{-1}$. In
the following, we set the current quark mass, $m_0$, to zero and 
assume the self-energy to take the (chirally invariant) form \be
\Sigma(\omega,{\bf k})= a(\omega,k) \gamma_{0}  + b(\omega,k)
\bf\hat k \cdot {\bf \gamma} \, , \label{Sigma-ab} \ee since scalar
and tensor contributions are suppressed due to chiral symmetry
restoration (above $T_c$), whereas pseudoscalar and axialvector
terms are absent due to parity invariance. The self-energy can be
further decomposed as (cf., \eg, Ref.~\cite{Kraemmer:2003gd}) \be
\gamma_0 \Sigma(\omega,{\bf k}) =
  \Sigma_+(\omega,k) \Lambda_+({\bf\hat k})
- \Sigma_-(\omega,k) \Lambda_-({\bf\hat k}) \ ,
\ee
where
\be
\Lambda_{\pm}({\bf\hat k})= \frac{1 \pm \gamma_0 {\bf\hat k \cdot
\gamma}}{2} \ 
\ee 
are projectors on quark states with chirality
equal ($\Lambda_+$) or opposite ($\Lambda_-$) to their helicity, and
$\Sigma_{\pm} = b(\omega,k) \pm a(\omega,k)$. The quark propagator
then follows as \be S(\omega,k)\gamma_0 =  \Delta_+(\omega,k)
\Lambda_+({\bf\hat k})
 + \Delta_-(\omega,k) \Lambda_-({\bf\hat k}) \,
\label{PropagatorFull} \ee with $\Delta_{\pm} = - (\omega \mp
(k+\Sigma_{\pm}))$. Since the potential extracted
from lQCD is independent of chirality, the self-energy
satisfies $\Sigma_{+} = - \Sigma_{-}$, that is $b(\omega,k) = 0$.
Recalling Eq.~(\ref{Sigma-ab}), this implies that the
non-perturbative interactions only contribute to a chirally
invariant (thermal) mass term for anti-/quarks. For the
``gluon-induced" self-energy, $\tilde\Sigma$ in Eq.~(\ref{Sigma0}),
we adopt a form suggested by the high-temperature hard-thermal loop
result, characterized by a mass term, $m$, in the pertinent
dispersion relation,
\be \omega_k = \sqrt{k^2+m^2} \ , \ee ignoring
possible imaginary parts. Our default value for $m$ is 0.1~GeV.

Let us now turn to the scattering equation (\ref{BS0}). As mentioned
above, we neglect the (virtual) antiparticle components in the quark
propagator and apply a 3-dimensional (3-D) reduction scheme to the
4-dimensional BS equation, facilitating its numerical evaluation
substantially. The resulting Lippmann-Schwinger equation takes the
form
\be
T_{a}(E;{\bf q^\prime},{\bf q}) = V_{a}({\bf q^\prime},{\bf q}) -
\int \frac{d^3k}{(2 \pi)^3} \ V_{a}({\bf q^\prime},{\bf k}) \
G_{q\bar q}(E; k) \ T_{a}(E;{\bf k},{\bf q}) \ [1-2f(\omega_k)]  \,
, \label{LS}
\ee
where $E$ denotes the center-of-mass ($CM$) energy
and ${\bf q}$ and ${\bf q}^\prime$ are the in- and outgoing (off-shell)
3-momenta in the $CM$ (as usual, the on-shell $T$-matrix is defined
by $q$=$q^\prime$ with $E=2\omega_q$ where $\omega_q$ is the
on-shell single-quark energy); $a=1,8$ labels color-singlet and
-octet channels, and \be f(\omega)\,=\,\frac{1}{e^{\omega/T}+1}\, \,
\ee is the Fermi-Dirac distribution. The explicit form of the
two-particle propagator, $G_{q\bar q} (E; k)$, depends on the 3-D
reduction scheme. Unless otherwise stated, we adopt the
Blankenbecler-Sugar (BbS)~\cite{Blankenbecler:1965gx} prescription
leading to
\be G_{q\bar q}(E; k)=
\frac{\omega_k}{\omega_k^2-E^2/4 + 2 i \omega_k
\Sigma_I(\omega_k,k)} \quad {\rm (BbS)}  \ , \label{GBbS}
\ee
but we
have checked that our results are very similar when employing the
Thompson scheme~\cite{Thompson:1970wt} with
\be G_{q\bar q} (E; k) =
\frac{1}{2} \frac{1}{\omega_k-E/2 + i \Sigma_I(\omega_k,k)} \quad
{\rm (Th)} \, \label{GTh} \,.
\ee
In both Eqs.~(\ref{GBbS}) and
(\ref{GTh}) $\omega_k$ denotes the on-shell quasiparticle dispersion
law, \ie, the solution of the equation
\be \omega_k =
\sqrt{k^2+m^2} + \Sigma_R(\omega_k,k) \, , \label{Quasiparticle}
\ee
with $\Sigma_R$ and $\Sigma_I$ the real and imaginary part of the
self-energy.
Finally, the potential figuring into Eq.~(\ref{LS}) follows from our
lQCD parametrization via Fourier transformation,
\be V_a({\bf
q^\prime},{\bf q}) = \int d^3r V_a(r) e^{i (\bf q - q^{\prime})\cdot
r} \, \label{potential}.
\ee
To solve Eq.~(\ref{LS}) it is
convenient to work in a partial-wave basis.
Expanding $T$-matrix and potential,
\bea
V_a({\bf q^\prime, q}) & =& 4
\pi \sum_l (2 l +1 ) \ V_{a,l}(q^\prime, q) \ P_l({\bf q^\prime
\cdot q}) \ ,
\\
T_a(E;{\bf q^\prime, q}) & =& 4 \pi \sum_l (2 l +1 ) \
T_{a,l}(E;q^\prime, q) \ P_l({\bf q^\prime \cdot q}) \ ,
\eea
allows to perform the angular integrations to yield
\be T_{a,l}(E;q^\prime, q) =
V_{a,l}( q^\prime, q) - \frac{2 }{\pi}\int k^2 d k \ V_{a,l}(
q^\prime, k) \ G_{q\bar q}(E; k) \ T_{a,l}(E;k,q) \ [1-2f(\omega_k)]
\ . \label{Tmatrix}
\ee
In the
present study we will constrain ourselves to $S$-wave channels,
deferring higher waves to future work. In Ref.~\cite{Shuryak:2004tx}
it was found that $P$-wave bound state formation is strongly
suppressed (in accordance with our own estimates).
Concerning spin-isospin channels, we recall that in the chirally
restored phase the
spectral functions of chiral partners (\eg, $\pi$-$\sigma$,
$\rho$-$a_1$) degenerate, which is also reflected in the spectral
functions extracted from lQCD~\cite{Asakawa:2002xj}.
Within the naive constituent quark model, $\pi$ and $\rho$ states
are $S$-wave $q$-$\bar q$ bound states, whereas $\sigma$ and $a_1$
are in a $P$-wave state. Interestingly, lQCD spectral functions
find an additional (approximate) degeneration of $\pi$ and $\rho$
states above $T_c$~\cite{Asakawa:2002xj,Karsch:2003jg}. In as far as an
interpretation of these objects as $q\bar q$ states applies, this
might be taken as an indication for a spin-symmetry much like in
heavy-quark effective theories. In view of these considerations, and
due to the fact that our lQCD-extracted potential is flavor-blind,
we will assume the color-singlet $S$-wave states to appear with a
spin-isospin degeneracy corresponding to $\pi$+$\rho$ states,
$d_{SI}=12$. Since the color-octet potential does not carry any
flavor-dependence either, the same factor will be applied to the
color-octet states.

With the $q$-$\bar q$ $T$-matrix at hand, we can proceed to
calculate the explicit expression for the quark self-energy due
to interactions with anti-quarks. Within the imaginary time
formalism the latter follows from closing the forward scattering
$T$-matrix with a thermal $\bar q$ propagator,
\be
\Sigma(z_v;p) =
\frac{d_{SI}}{12} \, d_a \int \frac{d^3p'}{(2\pi)^3} \ (-T) \
\sum\limits_{z_{\nu'}} T^a_{q \bar q}(z_\nu+z_{\nu'};{\bf p},{\bf
p}') \ D_{\bar q}(z_{\nu'},{\bf p}')
\ee
(here, $z_\nu\,=\,\pi i
(2\nu+1) T$ are fermionic Matsubara frequencies, $d_{1,8} = 1,8$ is
the color degeneracy factor and the factor $1/12$ represents the
average over the  $3 \times 2\times 2$
(color$\times$flavor$\times$spin) initial quark states). Using the
spectral representations of both $T$-matrix and $\bar q$ propagator
to perform the Matsubara sum, and after analytic continuation to the
real axis, the self-energy takes the form
 \be \Sigma_a(\omega;p)
= \frac{d_{SI}}{12} \, d_a \int \frac{d \omega'}{2 \pi} \int
\frac{dE}{\pi} \int \frac{d^3k}{(2 \pi)^3} \ A(\omega',k) \
\frac{f(\omega') + g(E)}{\omega+\omega'-E+i\eta} \ {\rm
Im}T_a(E;{\bf k}+{\bf p}) \, \label{Sigma} \ee with the Bose
distribution \be g(E)=\frac{1}{e^{E/T}-1} \, \ee and the quark
spectral function \be A(\omega,k)= \frac{- 2\, \Sigma_I(\omega,k)}
{(\omega -
\sqrt{k^2+m^2}-\Sigma_R(\omega,k))^2+\Sigma_I(\omega,k)^2}
\label{Spectral}\, . \ee To further simplify our task we assume in
the following a quasiparticle approximation for the spectral
function, \be A(\omega,k)\,=\, 2 \pi \delta(\omega-\omega_k) \,
,\label{SpectralFunction} \ee where $\omega_k$ is obtained from the
self-consistent solution of Eq.~(\ref{Quasiparticle}) (we will check
this approximation below). If we furthermore neglect the (weak)
energy dependence of $g(E)$ close to the pole of the principal value
integral in Eq.~(\ref{Sigma}), we can recover the real part of $T_{q
\bar q}$ to cast the self-energy in compact form, \be
\Sigma_a(\omega;p) = \frac{d_{SI}}{12} \, d_a \int \frac{k^2 dk\,
dx}{(2 \pi)^2} \  \ [f(\omega_k) + g(\omega+\omega_k)] \ T_{q \bar
q}^a(E) \,, \label{Sigma2} \ee
 where $x=\cos\theta$ (with
$\theta=\angle({\bf p},{\bf k})$) and the $CM$ energy of
the on-shell $T$-matrix is given by
\be E =
\sqrt{(\omega_k+\omega)^2-({\bf p}+ {\bf k})^2}\,.
\ee

\section{T-Matrix, Self-Energy  and Spectral Function}
\label{sec_res}
In this section we discuss the numerical solutions to the set of
equations (\ref{Quasiparticle}), (\ref{Tmatrix})  and
(\ref{Sigma2}). Self-consistency is achieved by iteration, starting
with the calculation of the $T$-matrix using a constant self-energy
in the first step. The self-energy is then calculated from
(\ref{Sigma2}) and used to solve the on-shell condition
(\ref{Quasiparticle}). The  pertinent quasiparticle dispersion-law
is then re-inserted into the $T$-matrix equation and the procedure
is iterated until $T$-matrix and self-energy converge (typically
within less than 10 iteration steps; we have also verified that the
final results are insensitive to the initial input value for the
self-energy).

\subsection{Quark-Antiquark $T$-matrix}
\label{subTmatrix}
The $T$-matrix equation (\ref{Tmatrix}) is solved using the matrix
inversion algorithm of Haftel and Tabakin~\cite{Haftel:1970} (after
discretizing the momentum integration). To assess the possible
formation of bound states, the $T$-matrix needs to be calculated
below the nominal $q$-$\bar q$ threshold,
$E_{thr}=2(m+\Sigma_R(E_{thr}/2,0))$. The potential does not depend
on the $CM$ energy $E$, and, due to its nonrelativistic character,
is only defined for real external 3-momenta $q$ and $q^\prime$. We
therefore define the subthreshold on-shell $T$-matrix by setting the
external momenta $q$=$q^\prime$=0. In the following we will refer to
a peak in the imaginary part of the $T$-matrix as a bound-state
(resonance) if the energy of the maximum is located below (above)
the quasiparticle threshold, $E_{thr}$.

\subsubsection{Charmonium Systems}
\begin{figure}[!th]
\includegraphics[height=2.1in,width=2.1in,angle=-90]{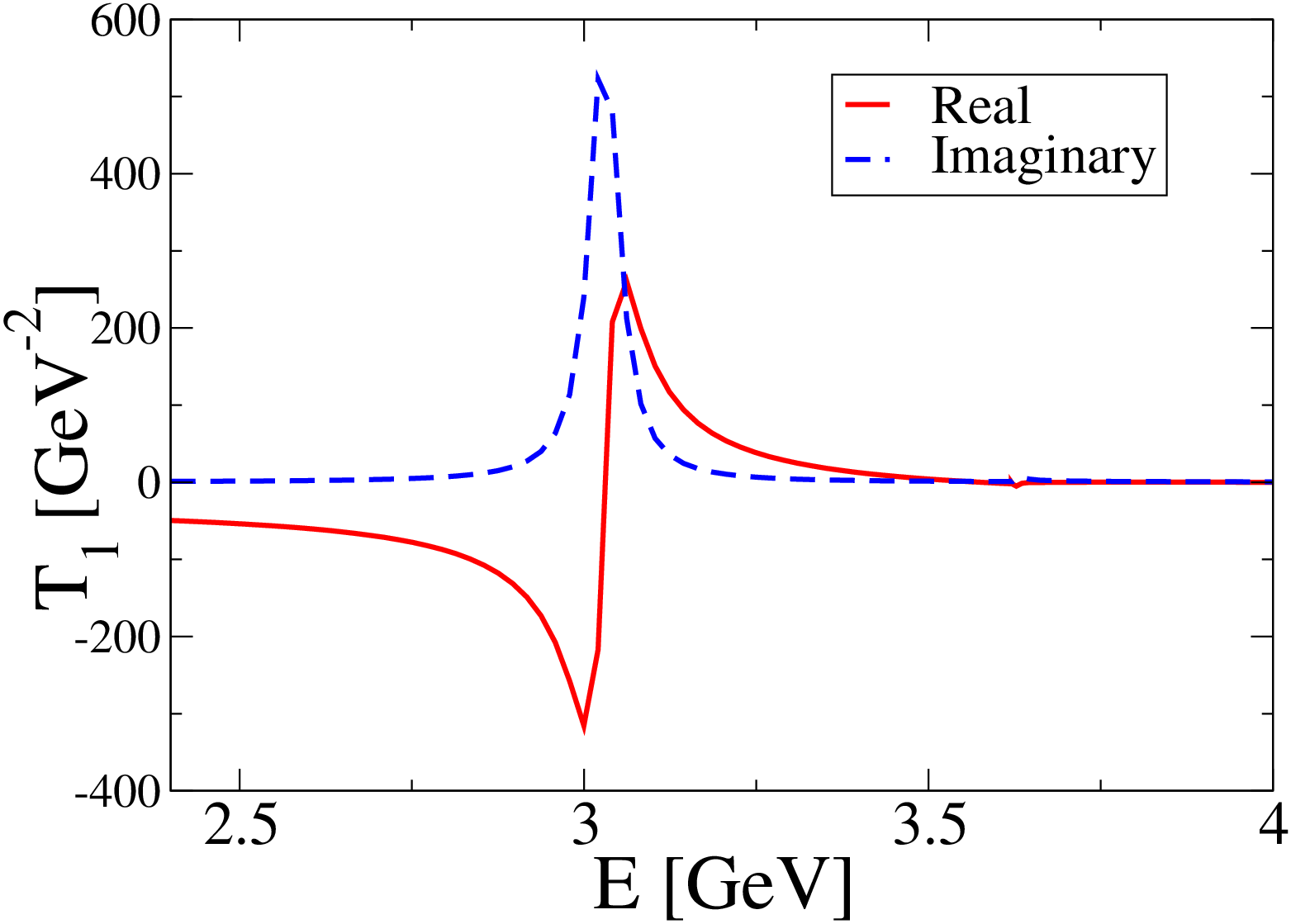}
\includegraphics[height=2.1in,width=2.1in,angle=-90]{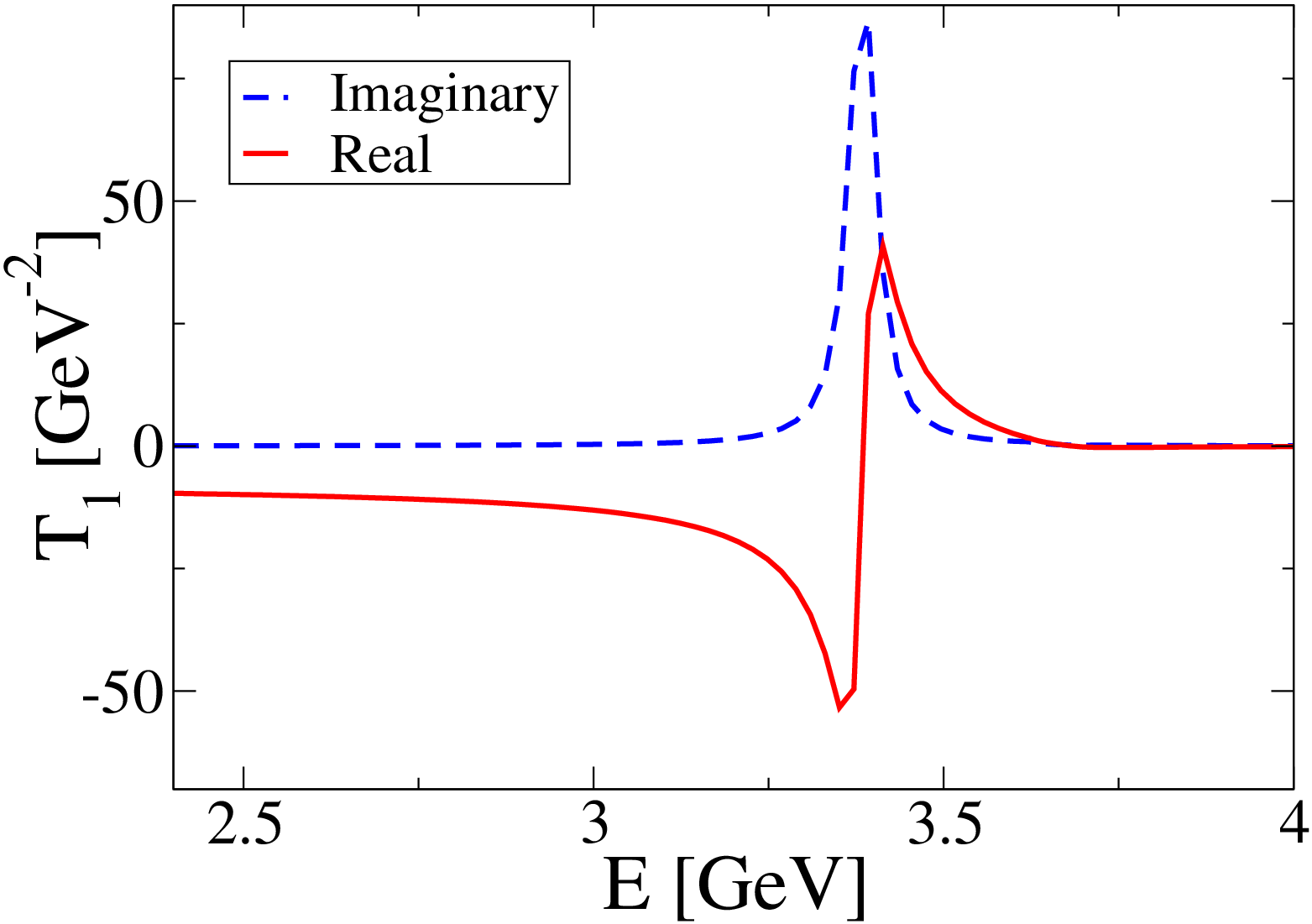}
\includegraphics[height=2.1in,width=2.1in,angle=-90]{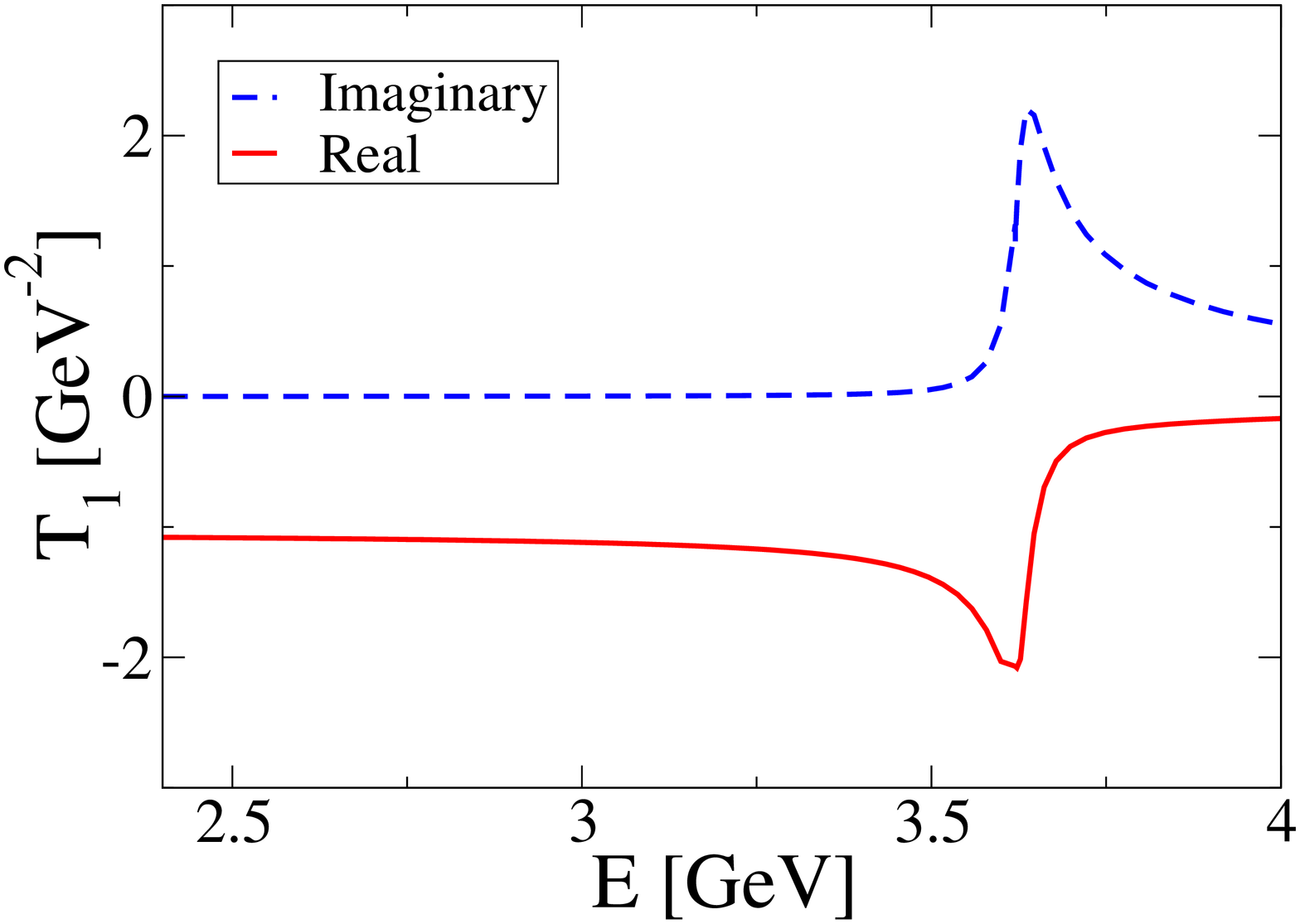}
\caption{Real (full red line) and imaginary part
(absolute value, dashed blue line) of $T$-matrix in the color-singlet 
channel for charmonium (with a charm-quark mass of $m=1.8$~GeV) at 
$T\,=\,1.2\,T_c$, $T\,=\,1.5\,T_c$, $T\,=\,2\,T_c$ (left, middle and 
right panel, respectively) as a function of $CM$ energy $E$, based
on our potential parametrization extracted from unquenched lQCD.}
\label{JPsi}
\end{figure}
\begin{figure}[!th]
\includegraphics[height=2.1in,width=2.1in,angle=-90]{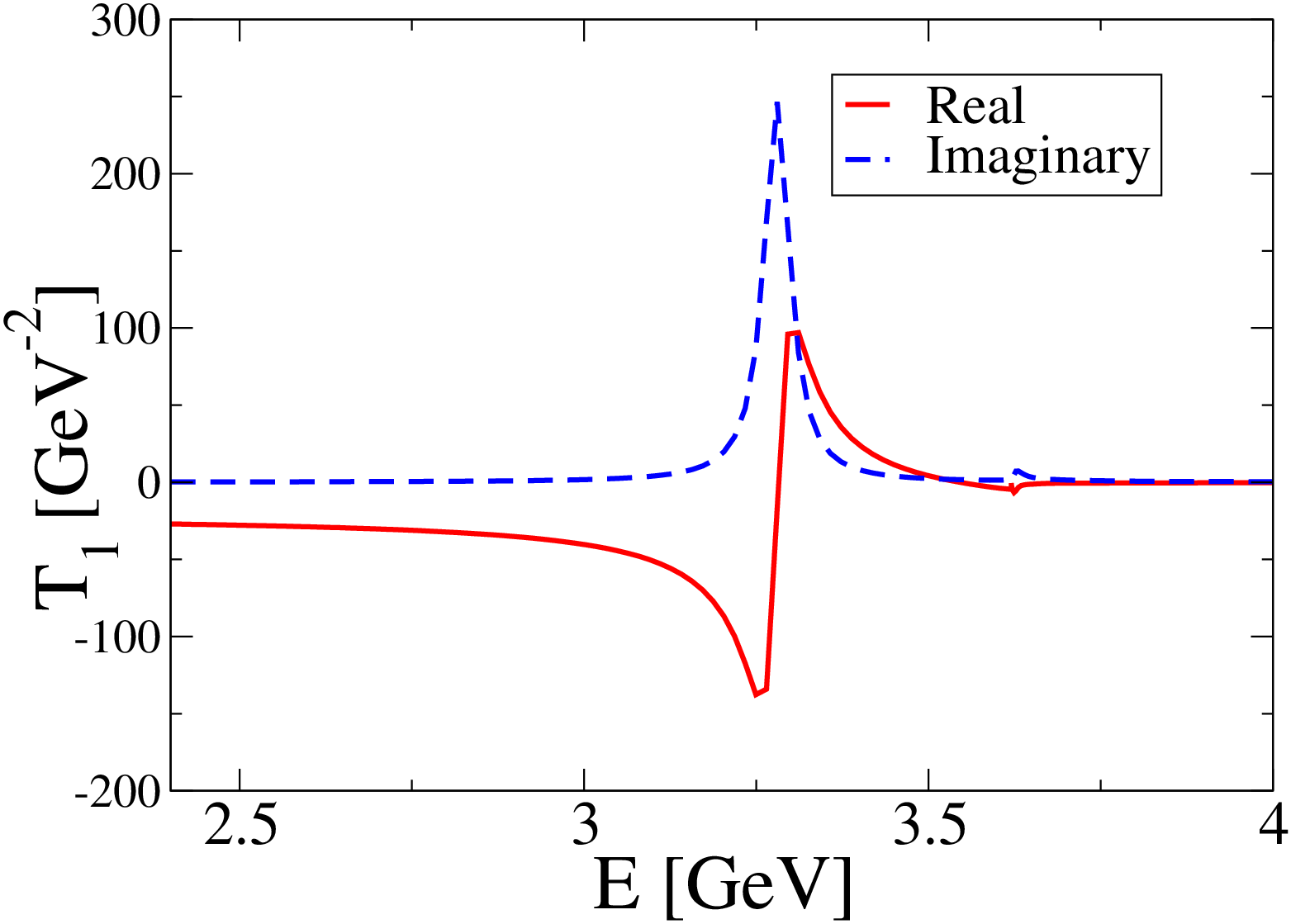}
\includegraphics[height=2.1in,width=2.1in,angle=-90]{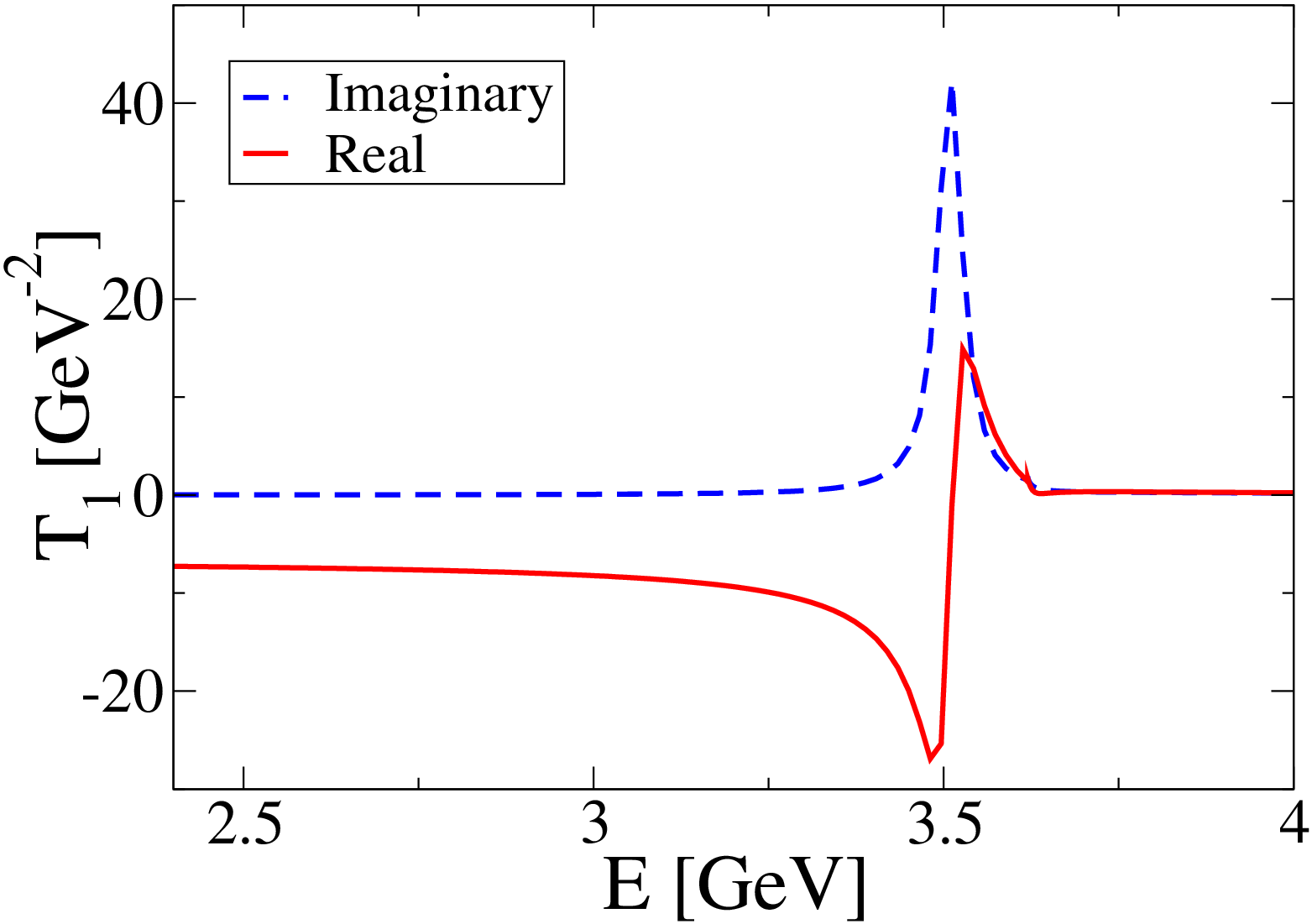}
\includegraphics[height=2.1in,width=2.1in,angle=-90]{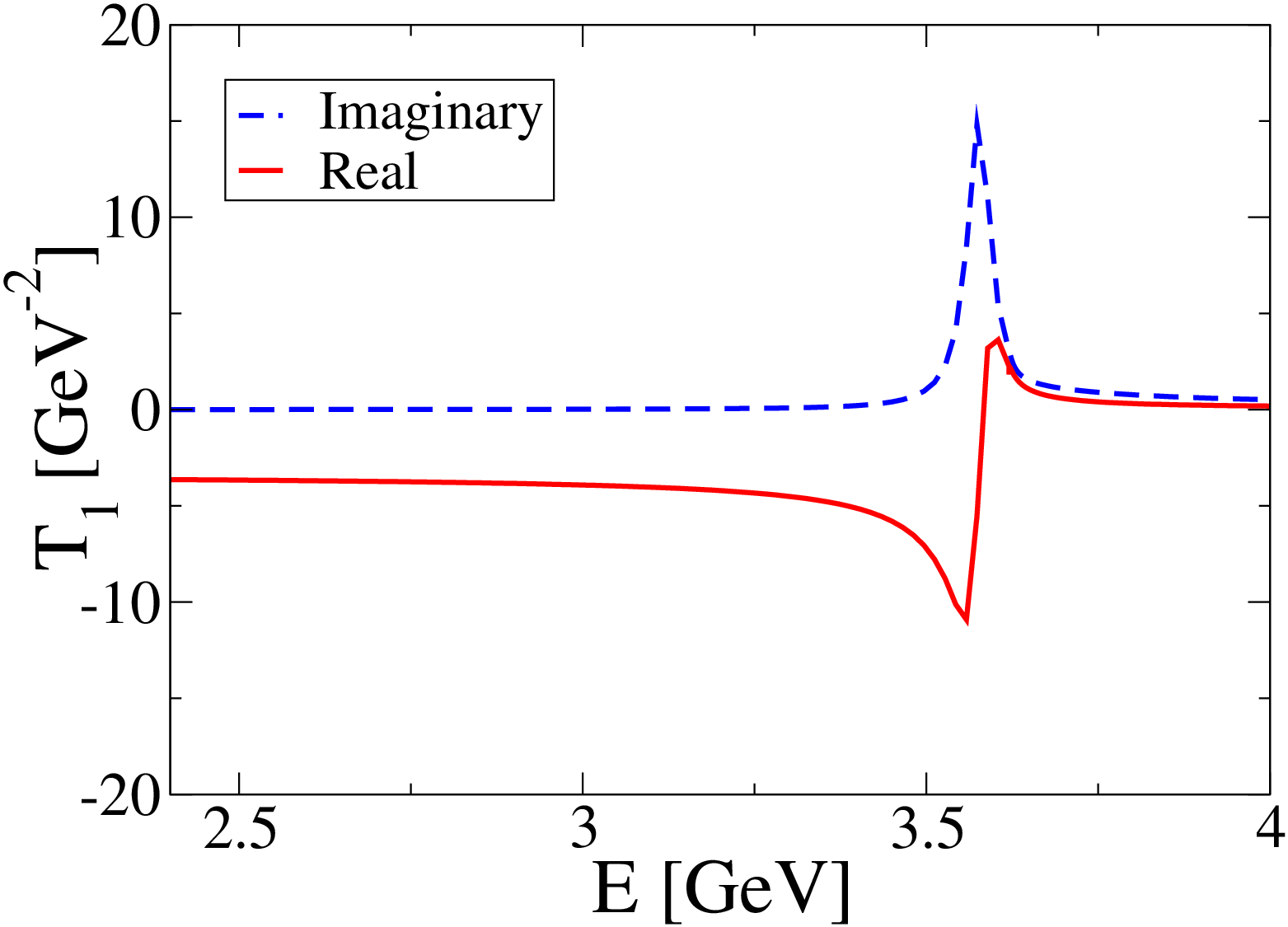}
\caption{Real (full red line) and imaginary part
(absolute value, dashed blue line) of $T$-matrix in the color-singlet
channel for charmonium (with a charm-quark mass of $m=1.8$~GeV) at
$T\,=\,1.2\,T_c$, $T\,=\,1.5\,T_c$, $T\,=\,2\,T_c$ (left, middle and
right panel, respectively) as a function of $CM$ energy $E$, based
on a potential~\cite{Wong:2004zr} extracted from quenched lQCD.}
\label{JPsi2}
\end{figure}

To check the reliability of the parametrization of the potential in
the singlet channel, and of the algorithm to compute the $T$-matrix,
we first apply our approach to the $c$-$\bar c$ (charmonium) sector
by using a (constant) quark mass of $m=$~1.8~GeV which approximately 
reproduces the vacuum $J/\psi$ mass at the lowest temperature (note 
that self-consistency does not play a role here since the thermal
abundance of $c$-quarks is strongly suppressed; for numerical
purposes, we used a fixed imaginary value for the self-energy,
$\Sigma_I =-10$~MeV, and $\Sigma_R=0$). The results are displayed
in Fig.~\ref{JPsi} for three different temperatures, $1.2\, T_c$
(left panel), $1.5\, T_c$ (middle panel) and $2\, T_c$ (right
panel). As the temperature increases the charmonium state moves up
in energy (reflecting a decreasing binding energy) reaching the
threshold ($E_{thr} = 3.6$~GeV) at $T \simeq 2\, T_c$ after which
the resonance peak essentially dissolves (also note that the
strength in the $T$-matrix is much reduced at 2~$T_c$ as
compared to the lower temperatures). This behavior is in
reasonable (qualitative) agreement with both lQCD
calculations~\cite{Asakawa:2003re} and effective potential models
using a Schr\"odinger equation~\cite{Wong:2004zr,Mocsy:2004bv}. 

To check the sensitivity to the underlying potential (recall the
discussion in Sec.~\ref{sec_latt} around Fig.~\ref{fig_comp}), we have 
repeated the calculations for the charmonium $T$-matrix in the singlet
channel using the (quenched-based) potential of Ref.~\cite{Wong:2004zr},
cf.~Fig.~\ref{JPsi2}. At the lower temperature of $1.2\, T_c$ the 
binding is significantly less pronounced (by about 0.25~GeV) as 
compared to our parametrization, as to be expected from the less 
attractive potential. At higher temperatures the agreement improves, 
and both potentials lead to a very similar temperature where the state 
crosses the $c$-$\bar c$ threshold (close to $2\, T_c$), with strongly 
reduced strength. This, in turn, is again in line with the 
Schr\"odinger-equation approach, in which the $c$-$\bar c$ system 
becomes unbound around $\sim$$2\, T_c$~\cite{Wong:2004zr}.
While the resonance at $2\, T_c$ appears to be rather
narrow, we recall that we did not include here (temperature dependent)
absorptive parts~\cite{vanHees:2004gq} and reduced masses for the 
$c$-quarks (nor inelastic charmonium reaction 
channels~\cite{Grandchamp:2003uw}), all of which are expected 
to increase the width of the charmonium states.

\subsubsection{Light-Quark Systems}
\begin{figure}[!th]
\includegraphics[height=2.1in,width=2.10in,angle=-90]{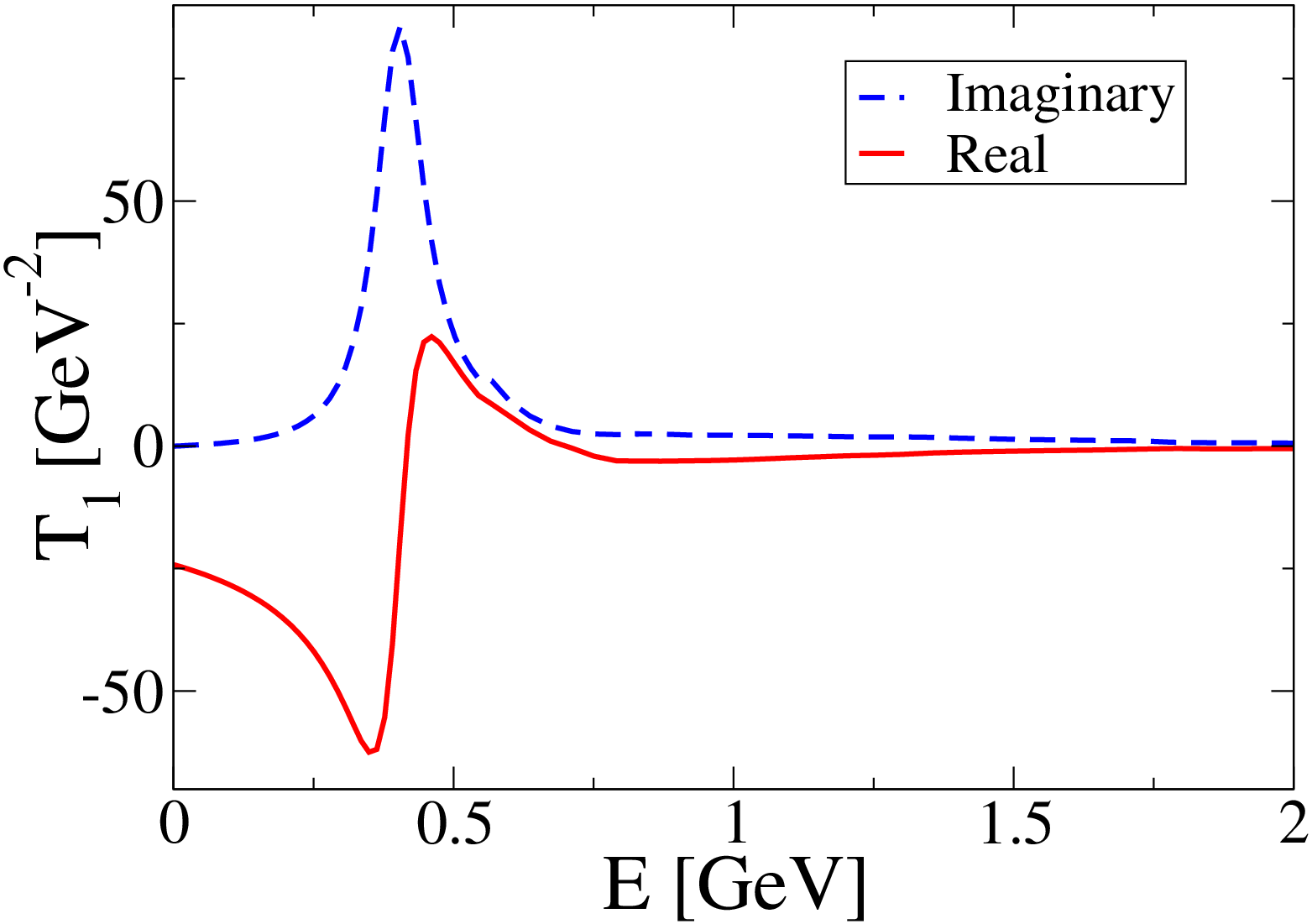}
\includegraphics[height=2.1in,width=2.10in,angle=-90]{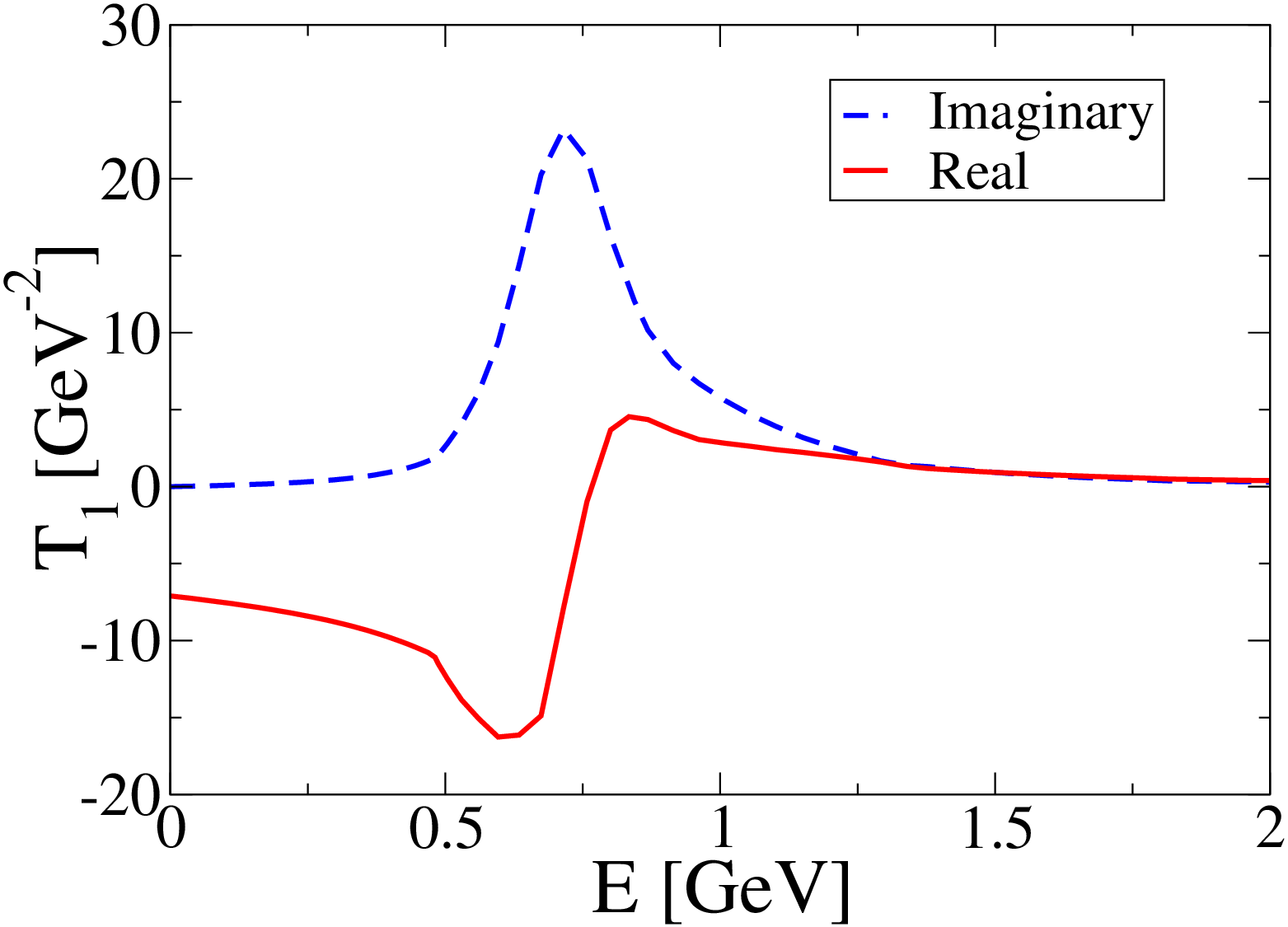}
\includegraphics[height=2.1in,width=2.10in,angle=-90]{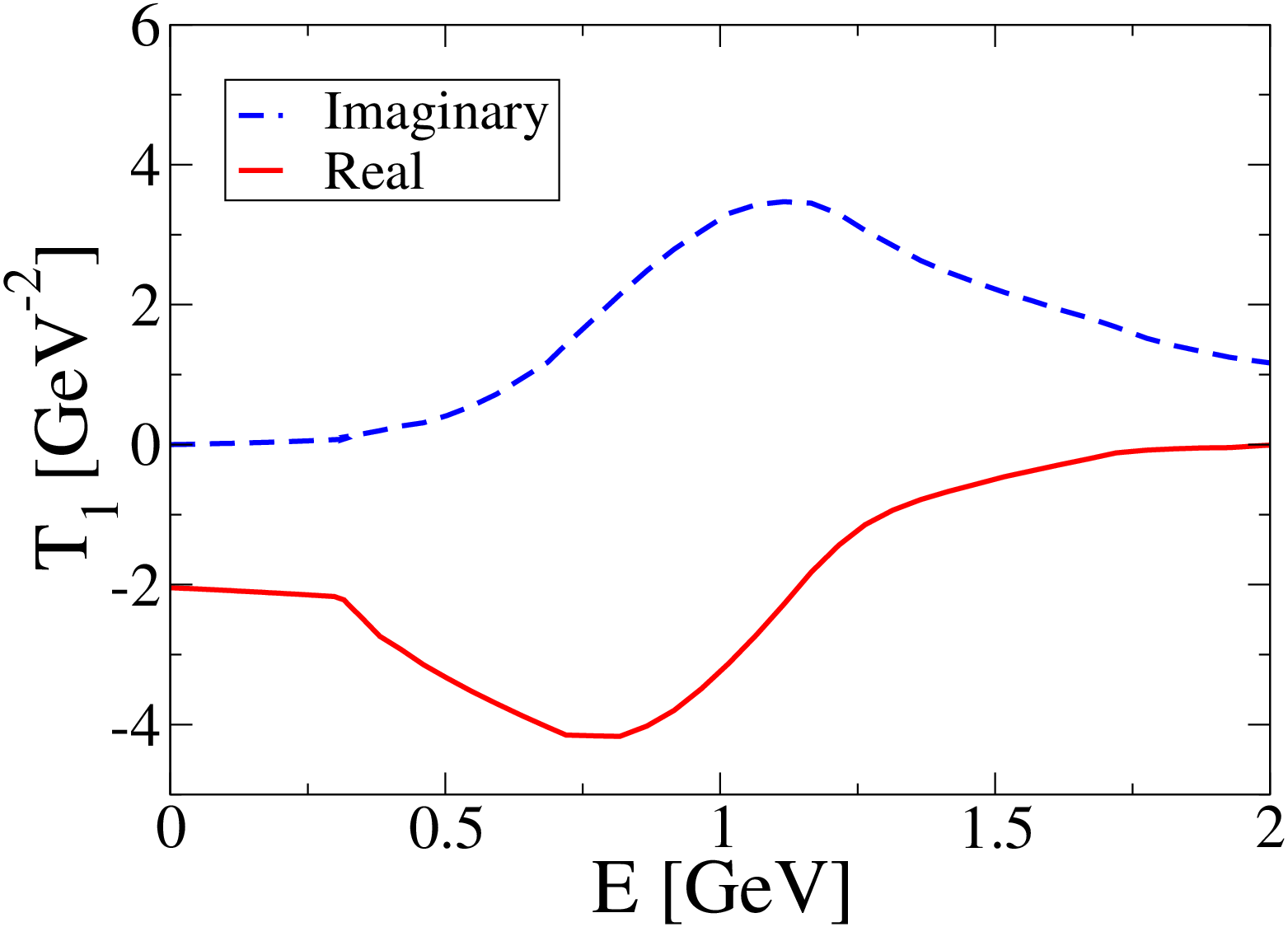}
\caption{Real (full red line) and (absolute value of the) imaginary part
(dashed blue line) of the light-quark (on-shell) $T$-matrix in the
color-singlet channel at temperatures $T\,=\,1.2\,T_c$, $T\,=\,1.5\,T_c$
and  $T\,=\,1.75\,T_c$ (left, middle and right panel, respectively)
as a function of the $q\bar q$ $CM$ energy $E$,
with a ``gluon-induced" quark-mass term $m=0.1$~GeV.}
\label{figTm1}
\end{figure}
\begin{figure}[!th]
\includegraphics[height=2.1in,width=2.10in,angle=-90]{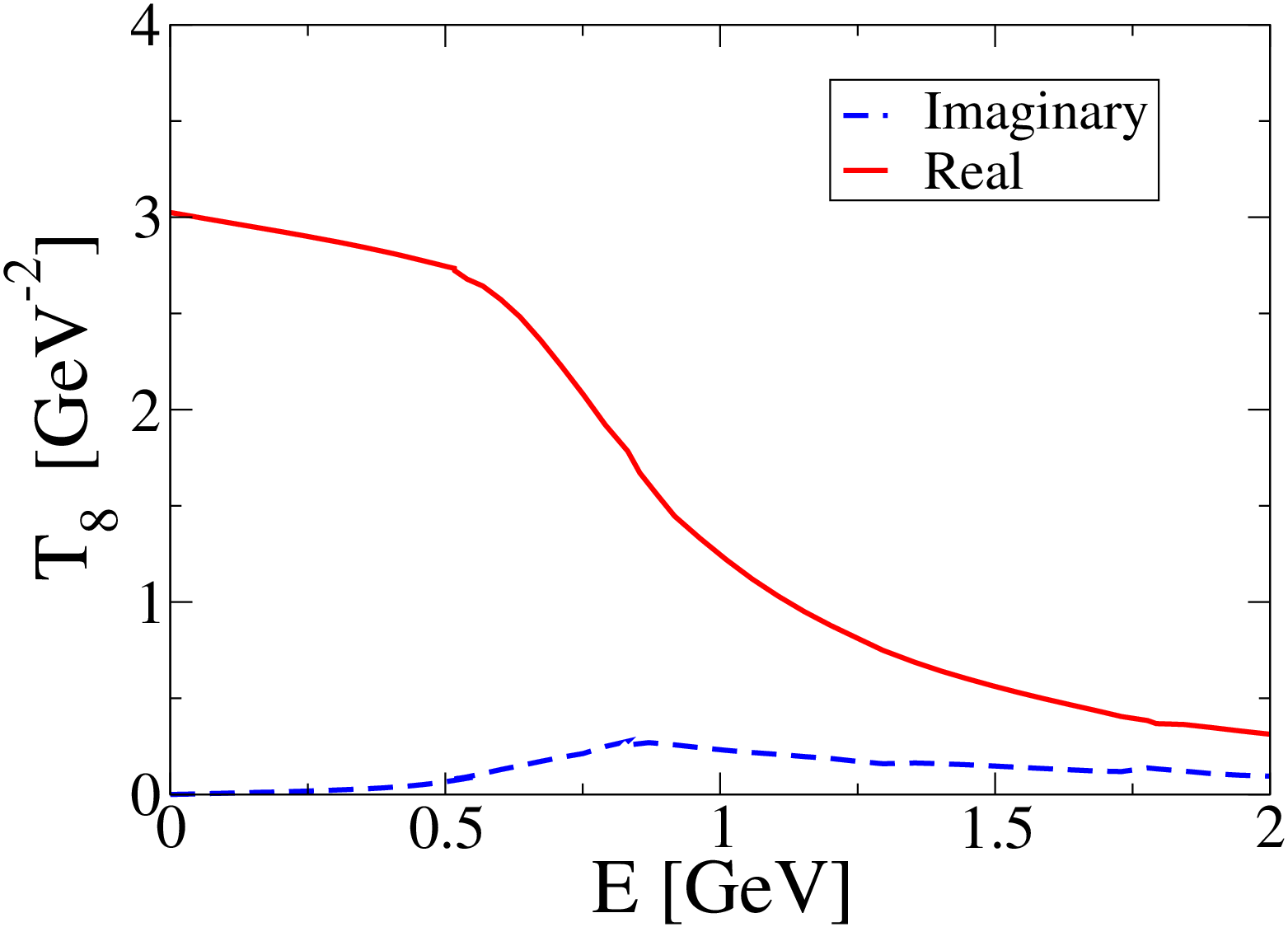}
\includegraphics[height=2.1in,width=2.10in,angle=-90]{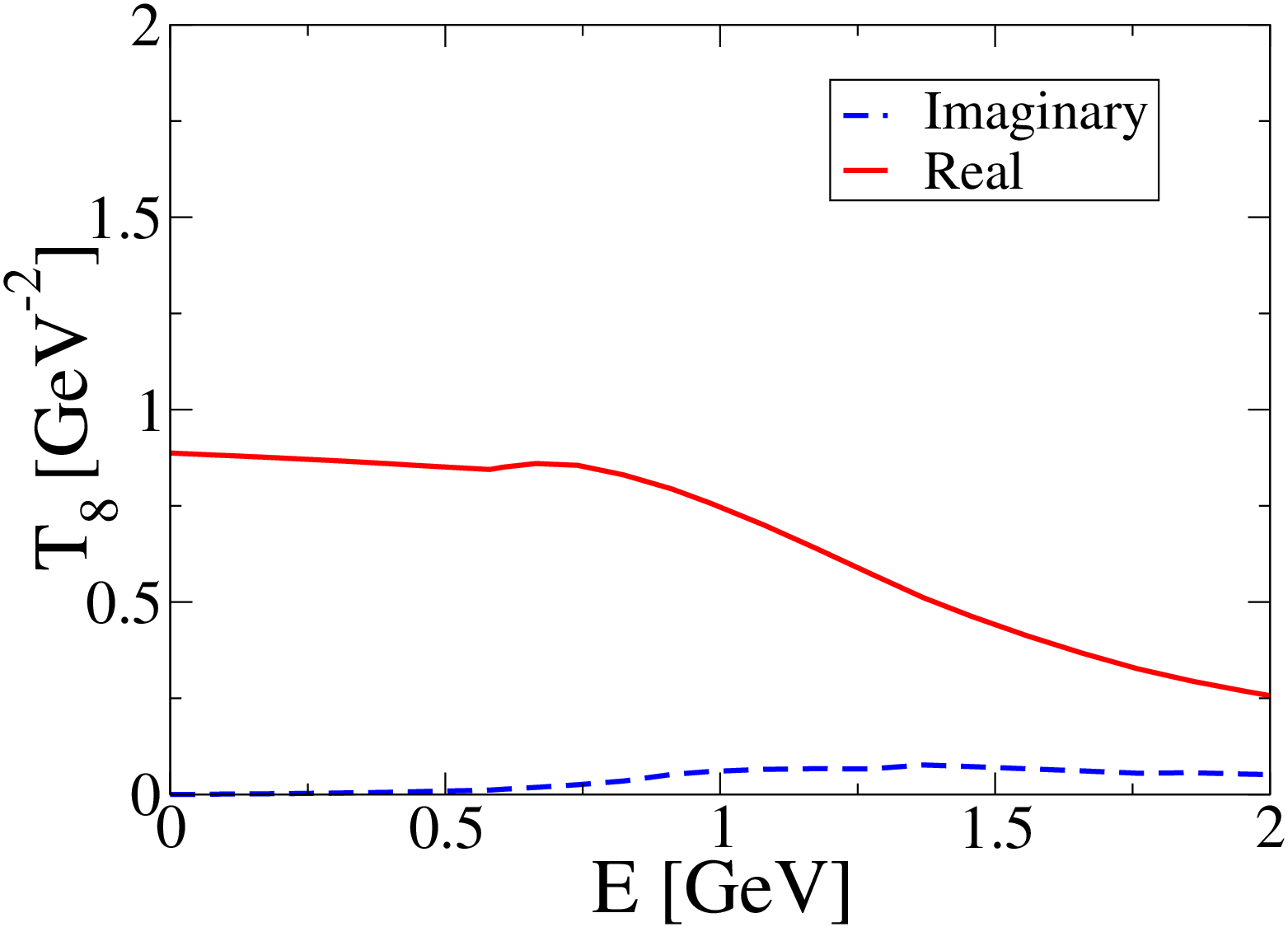}
\includegraphics[height=2.1in,width=2.10in,angle=-90]{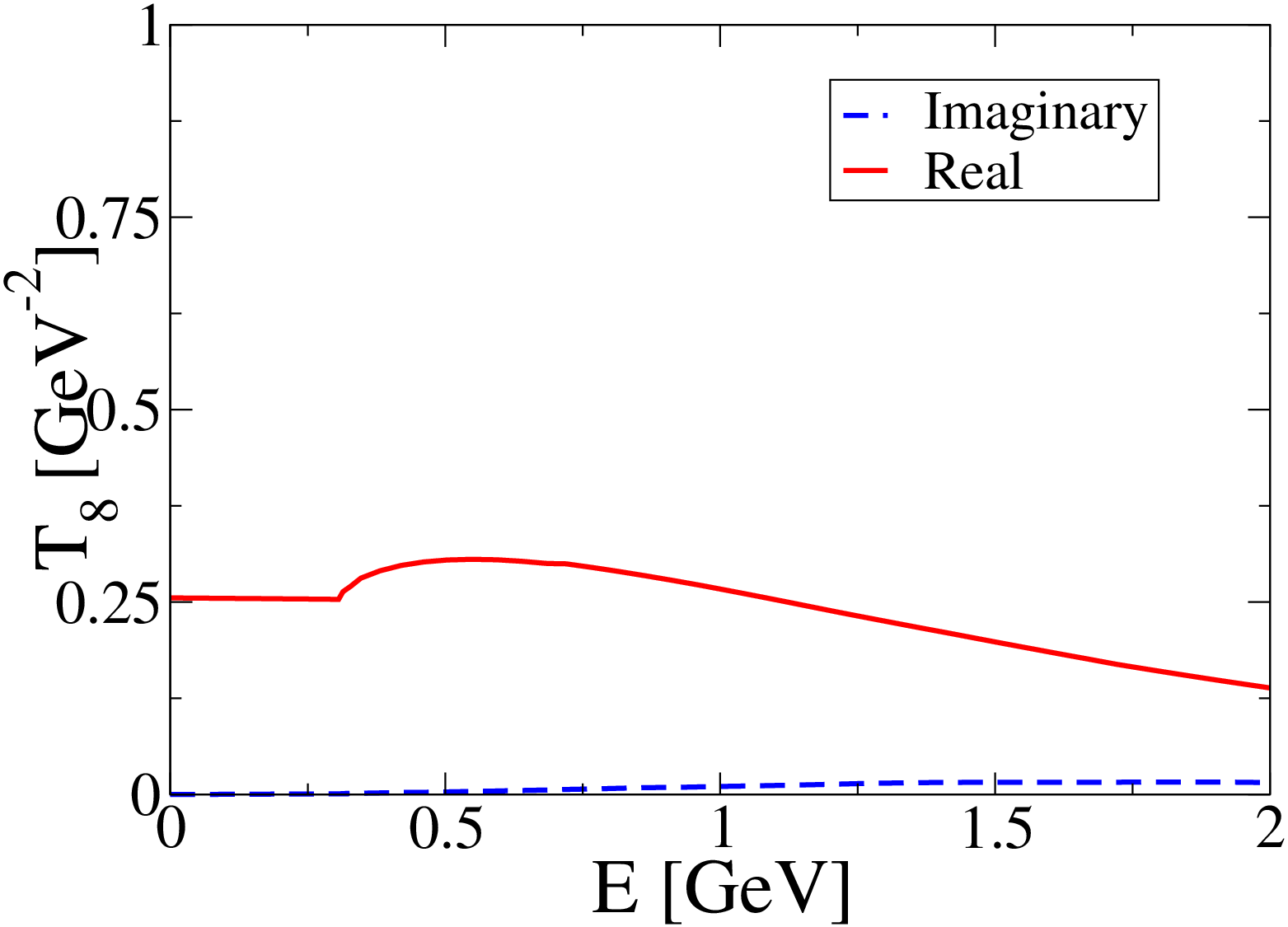}
\caption{Light-quark $T$-matrix in the color-octet channel vs.
$q\bar q$ $CM$ energy at $T\,=\,1.2\, T_c$, $T\,=\,1.5\, T_c$ and
$T\,=\,1.75\, T_c$ (left, middle and right panel, respectively)
with $m=0.1$~GeV. Solid (red) line: real part;
dashed (blue) line: imaginary part (absolute value).} 
\label{figTm2}
\end{figure}
Turning to the light-quark sector, the self-consistent results for real 
and imaginary part of the on-shell $T$-matrix for quasiparticles with a 
gluon-induced mass-term of $m=0.1$~GeV are summarized in 
Figs.~\ref{figTm1} and \ref{figTm2} for temperatures $T=1.2\, T_c$,
$T=1.5\, T_c$ and $T=1.75\, T_c$.

At $T=1.2\, T_c$ the color-singlet $T$-matrix  exhibits
a relatively narrow bound state located significantly below
the $q$-$\bar q$ threshold energy of $E_{thr}\simeq0.52$~GeV
(corresponding to twice the real part of the total quark self-energy
discussed below).
When increasing the temperature to $1.5\, T_c$, the state moves to
higher $CM$ energy above the threshold ($E_{thr}\simeq0.48$~GeV)
which, not surprisingly, is accompanied by a significant broadening.
Note also that the peak value is substantially reduced as compared
to the 1.2~$T_c$ case, substantially more than to be expected
from the broadening alone. We assign this behavior to the decrease in
the potential, cf.~right panel of Fig.~\ref{fig_lat}, reflecting
an overall reduction in interaction strength.
The trends in suppression, broadening and upward energy-shift
continue at $T=1.75\, T_c$
where the resonance has now essentially melted as indicated by a width
of almost 1~GeV, comparable to its mass.
These results may be put into context with computations of mesonic
spectral functions in (quenched) lattice QCD. For
(reasonably) light quarks~\cite{Asakawa:2002xj,Karsch:2003jg},
their main features above $T_c$ are a gradual increase of the peak
position (corresponding to the ``meson mass") with temperature
(roughly proportional to $T$), accompanied by a broadening. The
bound / resonance states depicted in Fig.~\ref{figTm1} approximately
share both of these features.

The $T$-matrix in the color-octet channel is displayed in
Fig.~\ref{figTm2} for the same set of temperatures. As to be expected
for a purely repulsive
potential, we find a smooth (non-resonant) dependence of both real
and imaginary part with $CM$ energy (with a substantial suppression
at higher $T$, as in the singlet case). The imaginary part is very
small, and also the real part appears to be small when compared to
the singlet channel. We recall, however, that the octet contribution
to the self-energy, Eq.~(\ref{Sigma2}), enters with a weight which
is by a factor of 8 larger than for the singlet one, rendering it an
important effect as will be seen below.

\subsection{Self-Energy}
\label{sec_self}
We proceed to the single-quark self-energies as calculated from the
interactions with antiquarks of the heat bath using the expression,
Eq.~(\ref{Sigma2}), based on the self-consistent $S$-wave
$q$-$\bar q$ $T$-matrices in the ``$\pi$" and ``$\rho$" channels as
obtained in the previous section. We recall that the real part of the
self-energy corresponds to a chirally invariant mass-term, whereas
its imaginary part determines the width of a quark (-quasiparticle)
according to $\Gamma=-2~$Im$ \Sigma$. Since we work at zero
quark-chemical potential, $\mu_q=0$, the same results hold for
antiquarks. We also recall that our on-shell approximation scheme
for the self-energy implies that the effects of bound states are not
captured by Eq.~(\ref{Sigma2}), since in the integration over the
$T$-matrix only energies above the $q$-$\bar q$ threshold, $E_{thr}$,
contribute.

\begin{figure}[th]
\includegraphics[height=2.1in,width=2.10in,angle=-90]{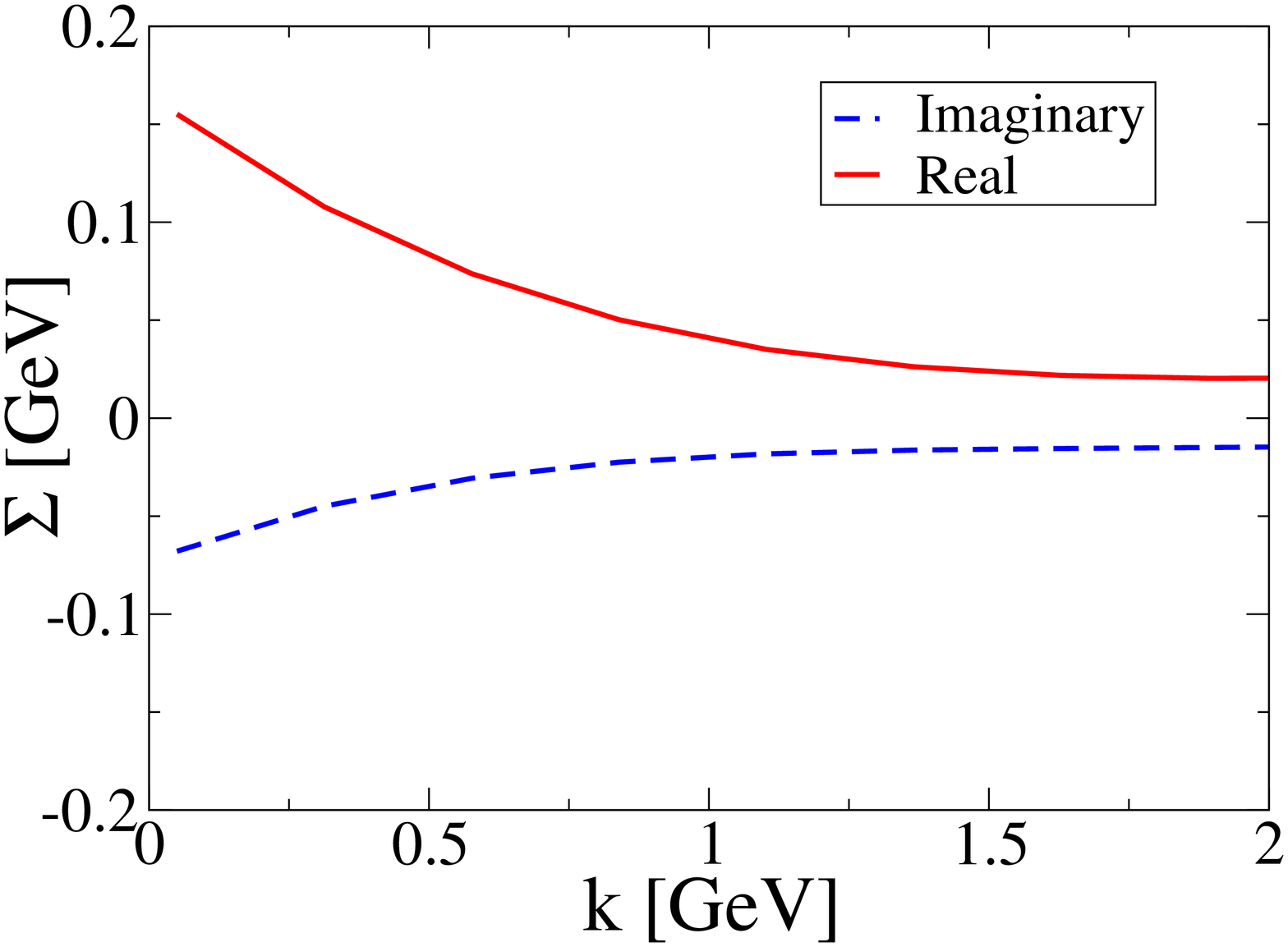}
\includegraphics[height=2.1in,width=2.10in,angle=-90]{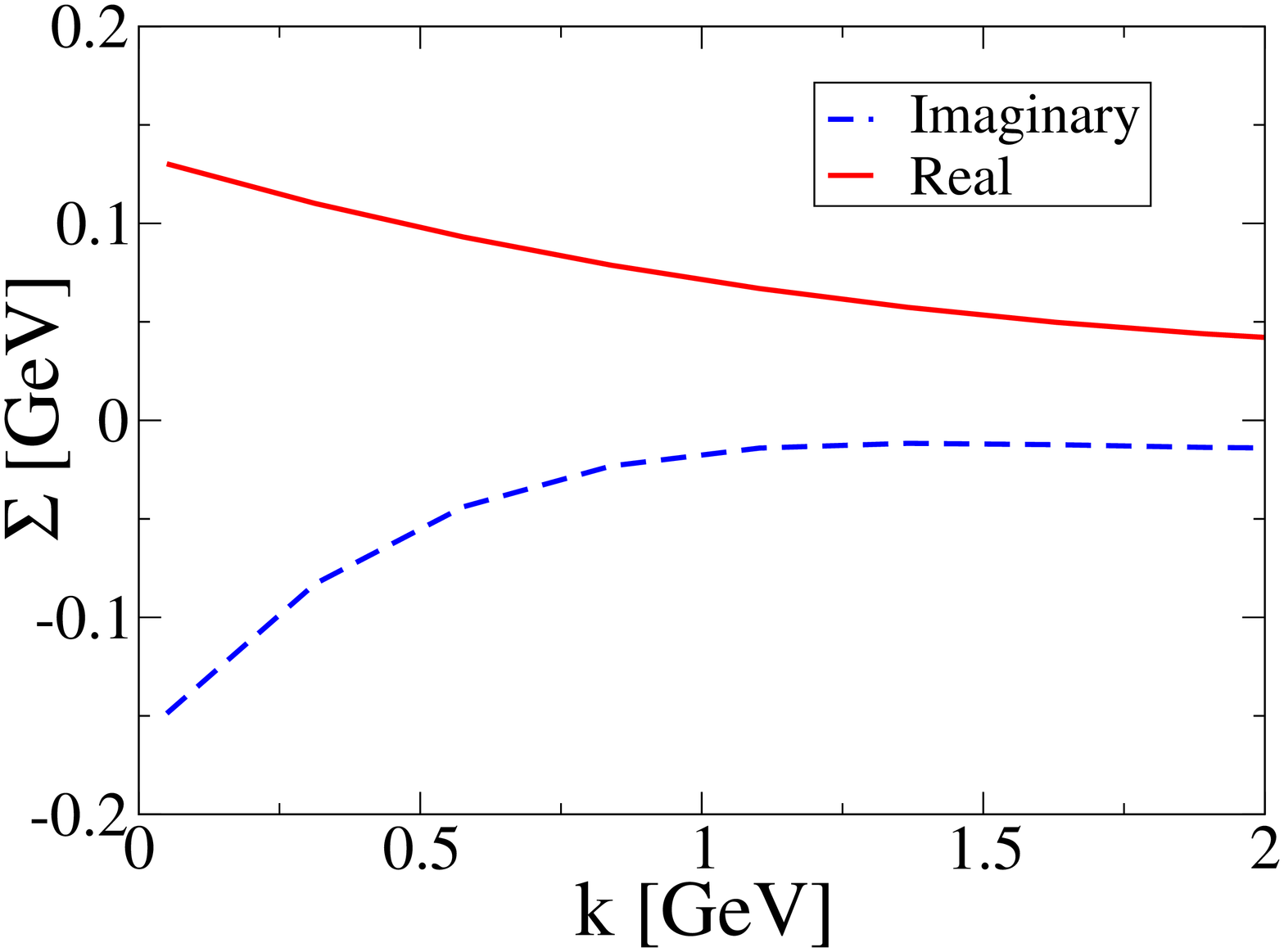}
\includegraphics[height=2.1in,width=2.10in,angle=-90]{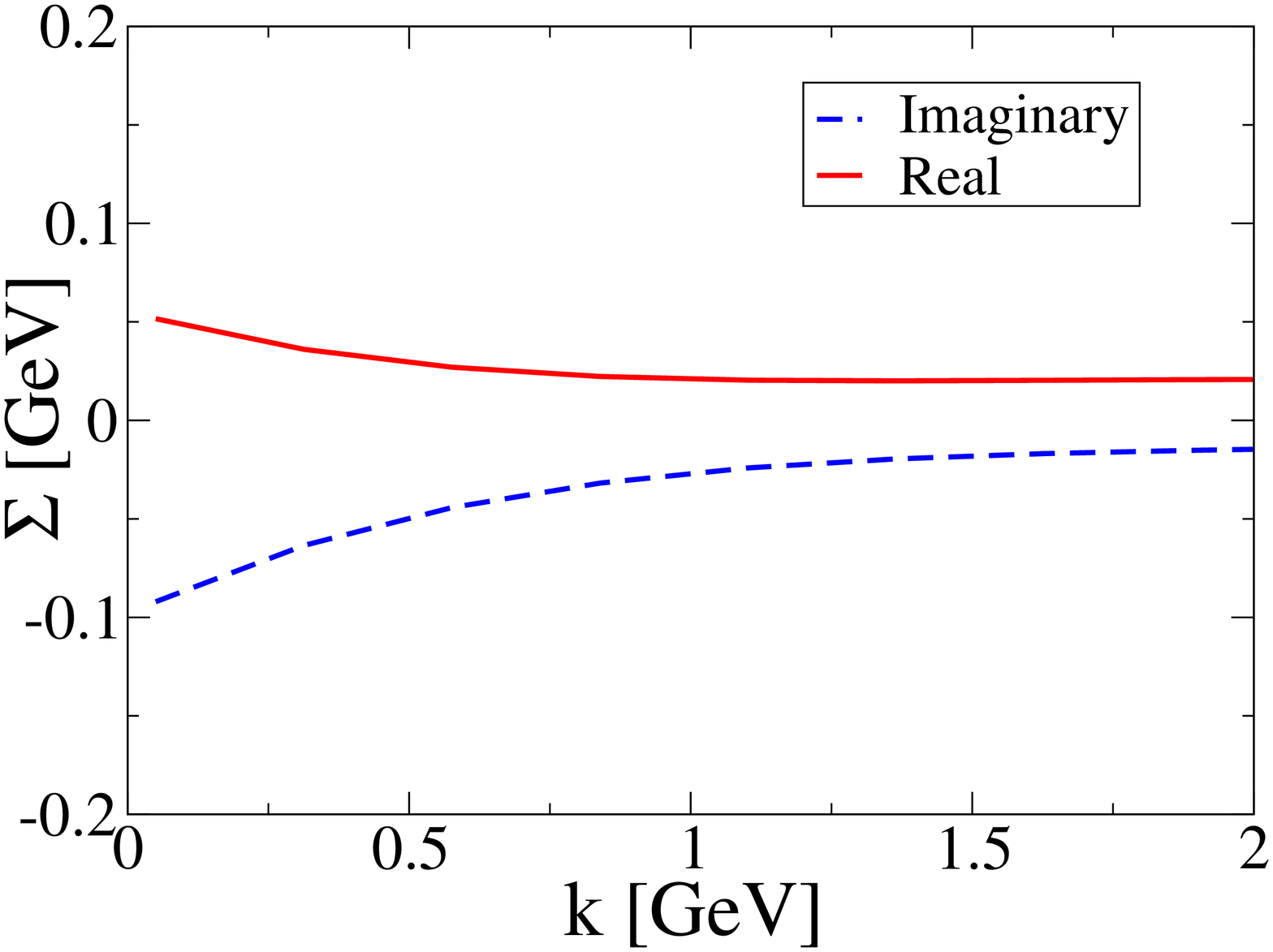}
\caption{Real (solid line, red) and imaginary (dashed line, blue)
part of the on-shell quark self-energy as a function of 3-momentum
at temperatures $T\,=\,1.2\, T_c$, $T\,=\,1.5\, T_c$ and
$T\,=\,1.75\, T_c$ (left, middle and right panel, respectively) with
$m=0.1$~GeV.} \label{figSelf1}
\end{figure}
In Fig.~\ref{figSelf1} the on-shell self-energy is displayed for the
same selection of temperatures as in the previous section. Both real
and imaginary part are smooth functions of the quark 3-momentum with
maximal values at $k=0$. Note that real part is positive, implying
that the repulsive contribution from the octet channels overcomes
the attractive singlet channels. The imaginary part (width), on the
other hand, chiefly arises due to resonant scattering in the singlet
channel.

\begin{figure}[!th]
\includegraphics[height=2.1in,width=2.10in,angle=-90]{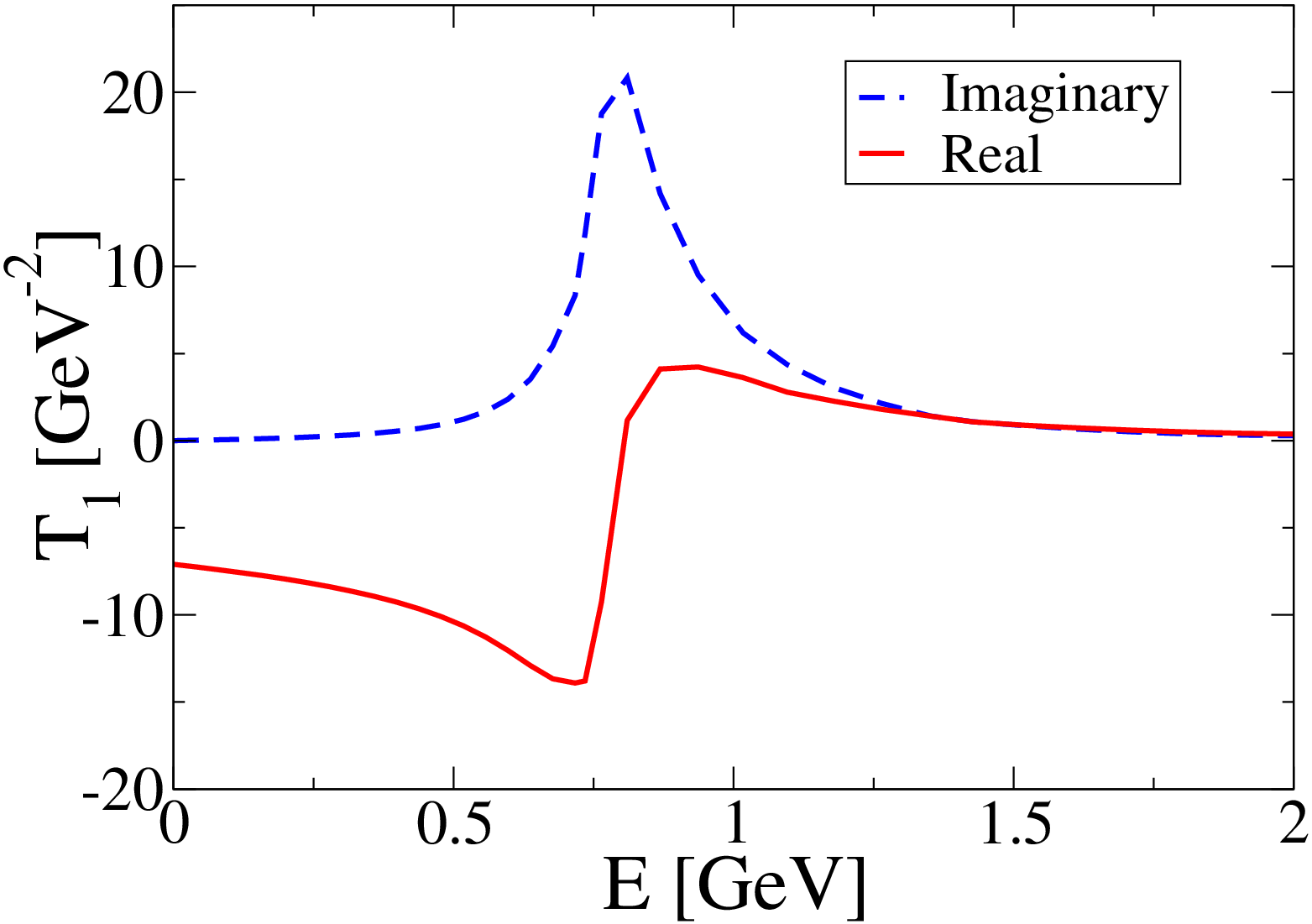}
\includegraphics[height=2.1in,width=2.10in,angle=-90]{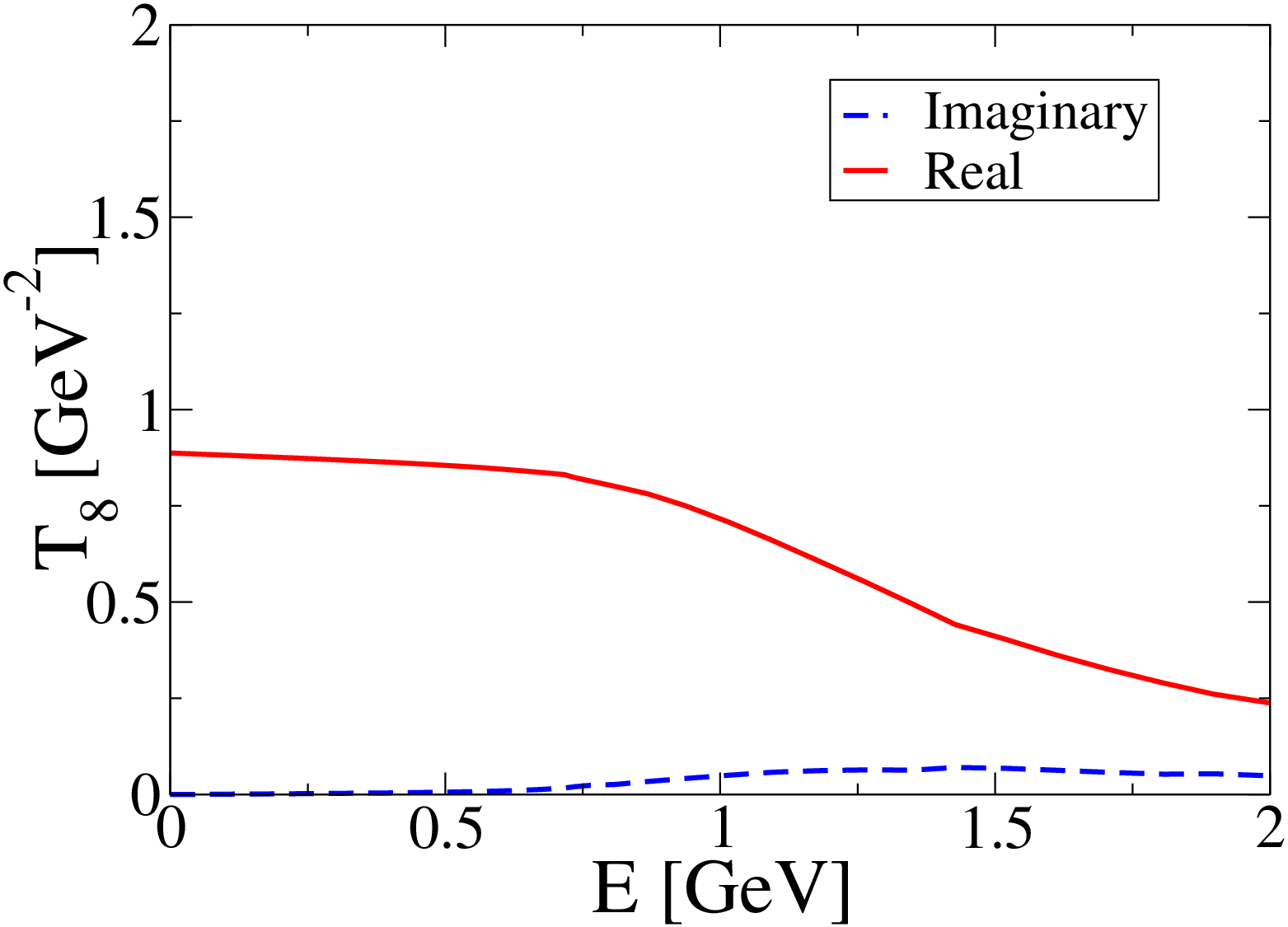}
\includegraphics[height=2.1in,width=2.10in,angle=-90]{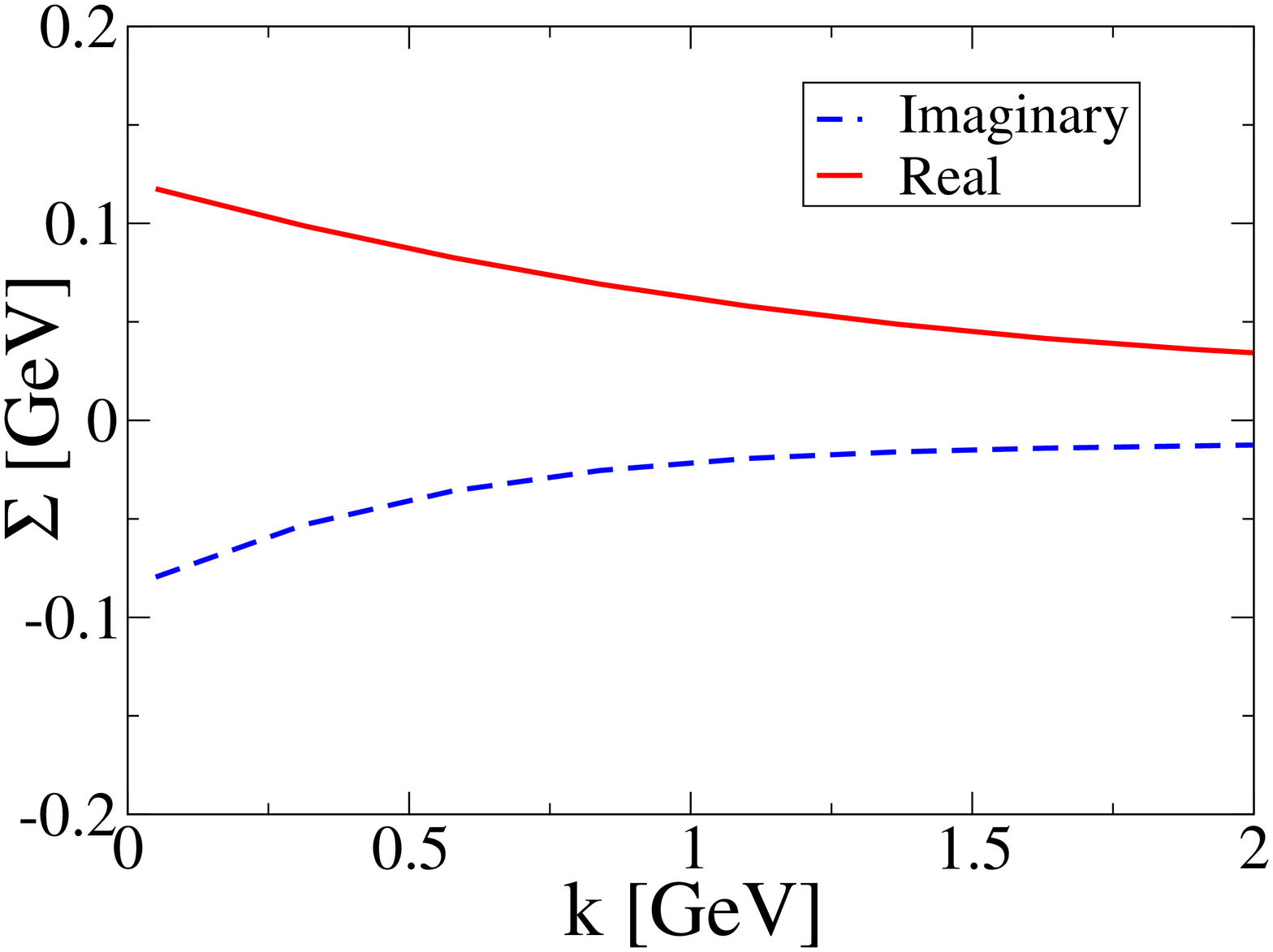}
\caption{Real (full line, red) and imaginary (dashed line, blue)
parts of the $T$-matrix  in the color-singlet channel (left panel),
color-octet channel (central panel) and corresponding
(singlet+octet)  self-energy (right panel) at a temperature
$T\,=\,1.5\,T_c$ using a ``gluon-induced" mass term of
$m=0.25$~GeV.} \label{figTandsigma}
\end{figure}

More quantitatively, in the temperature regime 1.2-1.5$\,T_c$,
the nonperturbative contribution to the thermal quark mass reaches
values of around 150~MeV at small momenta,  decreasing to
$\sim$50~MeV at 1.75~$T_c$. With the underlying ``gluon-induced"
mass term of $m=100$~MeV, the total thermal mass, $m + \Sigma_R$,
amounts to 150-250~MeV. This is smaller than effective
(perturbative) thermal quark masses required in phenomenological
fits to the QGP EoS of
lQCD~\cite{Levai:1997yx,Schneider:2001nf,Peshier:2002ww,Blaizot:2003tw}.
To improve
upon this, we have performed self-consistent calculations with a
gluon-induced mass term of $m=250$~MeV. It turns out that, at given
temperature,  the ``mesonic" states are slightly stronger bound,
but in general the behavior of $T$-matrix
and pertinent self-energy are quite similar to those obtained with
$m=100$~MeV. {\it E.g.}, for $T=1.5~T_c$ (cf.~Fig.~\ref{figTandsigma}),
the resonance structure is right at threshold, the quark width
reaches almost 200~MeV, and the combined real part at low momenta
amounts to a quark mass of $m+\Sigma_R \simeq 350$~MeV.

An important aspect of our results are the rather large imaginary
parts of the quark self-energy, translating into widths of about
200~MeV at low momenta for temperatures around 1.5~$T_c$. As
mentioned above, the width is almost entirely generated by the
resonant scattering in the singlet channel; this is nicely
illustrated by the significant increase in Im$\Sigma$ when going
from 1.2 to 1.5~$T_c$ (cf. left and middle panel in
Fig.~\ref{figSelf1}), during which the state in the $T$-matrix moves
from below to above threshold (cf. left and middle panel in
Fig.~\ref{figTm1}), \ie, converts from bound state to
resonance\footnote{We recall that bound states are not accessible in
on-shell 2~$\to$~2 scattering; even if a resonance is close to
threshold it does not contribute effectively to rescattering
processes if the average thermal energy of particles from the heat
bath is significant. The contribution of bound states to the
self-energy can be included rather by going beyond the quasiparticle
approximation, \ie,  evaluating Eq.~(\ref{Sigma}) with the off-shell
spectral function, Eq.~(\ref{Spectral}).}. The magnitude of the
quark widths is quite comparable to the thermal masses,
qualitatively supporting the notion that the QGP could be in a
liquid-like regime~\cite{Peshier:2005pp,Thoma:2005uv}. Even at the
highest considered temperature of 1.75~$T_c $, and at typical
thermal momenta ($k\simeq 3T\simeq0.9$~GeV), the quark width due to
scattering off antiquarks is between 50 and 100~MeV. This does
neither include $P$-wave interactions, nor strange antiquarks, nor
any contributions from scattering off quarks or gluons.

Finally let us come back to the uncertainty associated with the 
interaction in the octet channel related to the perturbative
ansatz, Eq.~(\ref{V8}). If the coeffiecient in Eq.~(\ref{V8}) is 
increased (decreased) by a factor 2 (for $T=1.5\,T_c$ and $m=0.25$~GeV), 
the imaginary part of the self-energy barely changes, 
whereas its real part increases (decreases) by about a 40\%.

\subsection{Quark Spectral Functions and Normalization Condition}
\begin{figure}[!th]
\includegraphics[height=3.in,width=3.0in,angle=-90]{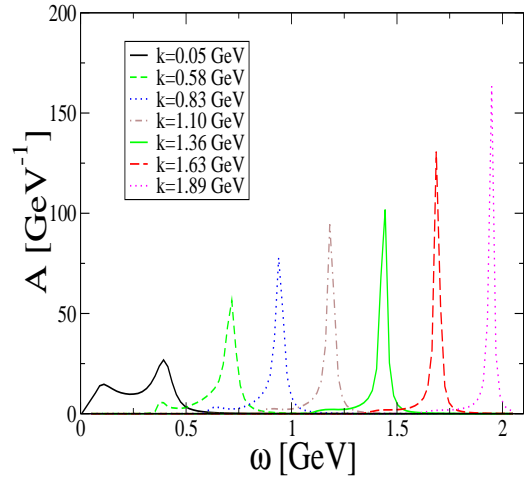}
\caption{Off-shell spectral function $A(\omega,k)$ as given by
Eq.~(\ref{Spectral}) vs. quark energy for different
values of the quark momentum at $T\,=\,1.5\, T_c$ and for
$m=0.25$~GeV. } \label{fig_2}
\end{figure}
To better elucidate the validity of the quasiparticle approximation,
Eq.~(\ref{SpectralFunction}), we compute the off-shell
real and imaginary parts of the self-energy using Eq.~(\ref{Sigma2})
and obtain the pertinent quark spectral function $A(\omega,k)$
from Eq.~(\ref{Spectral}).
In Fig.~\ref{fig_2} we depict $A(\omega,k)$ as a function of quark 
energy, $\omega$, for various fixed momenta at a temperature of
$T=1.5~T_c$. On the one hand, this
reiterates the large effect of the width for low momenta and calls
for an off-shell treatment to improve the reliability of our results
in the (sub-) threshold region of the $T$-matrix. On the other hand,
for larger momenta (including typical thermal momenta) the
quasiparticle approximation as applied in our calculations
appears to be reasonably well justified.

As another check of our approximations we have evaluated the norm
of the quark spectral functions defined by
\be I(k)\,=\,\int\frac{d
\omega}{2 \pi} A(\omega,k) \label{unitary}\, .
\ee
The unitarity
condition for $A(\omega,k)$ requires $I(k)=1$ for each
momentum $k$.  This relation is rather well satisfied,
$I(k)\geq 94~\%$, for all momenta considered in Fig.~\ref{fig_2}.

\section{Conclusions and Outlook}
\label{sec_concl}
In the present article we have set up a self-consistent many-body
scheme of Brueckner-type to assess nonpertubative properties of
(anti-) quarks and mesonic composites in a Quark-Gluon Plasma at
temperatures $T\simeq$~1.2-2$\,T_c$. Our key ingredient to describe the
$q$-$\bar q$ interaction in the QGP was a driving kernel (potential)
extracted from unquenched finite-$T$ lattice QCD calculations for
the free energy of a heavy-quark pair, supplemented with corrections
for relativistic motion. Our main objective was to go beyond earlier
applications to bound states by solving the scattering problem
thereby accounting for absorptive effects (finite imaginary parts).
The self-consistent set of single-quark Dyson and two-body
scattering equations has been solved by numerical iteration
employing a nonrelativistic reduction of the Bethe-Salpeter equation
in connection with a quasiparticle approximation for the quark
propagators. One of our main new findings is that the lQCD
potentials (dynamically) generate $S$-wave resonance states {\em
above} the $q$-$\bar q$ threshold up to temperatures of $\sim2~T_c$.
These resonances (assumed to occur with a degeneracy corresponding
to ``$\pi$"- and ``$\rho$"-mesons), in turn, play a key role in
inducing large quark scattering rates (=imaginary parts of the quark
self-energy) as indicated by single-particle widths of
$\Gamma\simeq$~200~MeV at temperatures around 1.5~$T_c$.  At the
same time, significant (positive) real parts arise from repulsive
interactions in the color-octet channel entailing thermal masses of
up to $\sim$150~MeV. We expect that additional contributions to the
quark mass of $\sim$250~MeV (induced, \eg, by interactions with
thermal gluons as parametrized in quasiparticle models) will be
necessary to account for the QGP EoS computed in lattice QCD.
Nevertheless, especially at low momenta, the quark widths are
comparable to the thermal masses, which could be indicative for
liquid-like properties of the QGP at moderate temperatures.

Our analysis suggests several directions for future work. First, the
accuracy of our approximations should be scrutinized. This includes
improving upon the quasiparticle approximation of the quark spectral
function by implementing its off-shell (energy-) dependence (as,
\eg, carried out in Ref.~\cite{Rapp:1995fv} for a hot pion gas),
most notably at low energies to incorporate bound-state
contributions to the quark self-energy. The scattering equation
ought to be extended to finite total 3-momentum of the mesonic
composites. Even though we expect the $q$-$\bar q$ channel to
constitute a major part of the in-medium interaction, a more
complete treatment including $q$-$q$ and $q$-$g$ channels is
desirable. It is also conceivable that processes of the type $q \bar
q \to M g$ (inverse gluon-dissociation; $M$: mesonic state) could be
significant, as they render bound states accessible in (on-shell)
2-body scattering. In a broader context, the underlying EoS of the
interacting system needs to be investigated, which is obviously not
an easy task. On the phenomenological side, to address the problem
of early equilibration at RHIC, it will be of great interest to
calculate the thermal equilibration timescales for (anti-) quarks
based on the resonant scattering amplitudes found here (\eg, within
a Fokker-Planck equation). The elastic scattering rates of around
1/(fm)/$c$ as found in this work, together with the isotropic
angular dependence inherent in $S$-wave rescattering, look
promising. For gluons the situation could be more involved since,
besides bound states as suggested in Ref.~\cite{Shuryak:2004tx},
other thermalization mechanisms might be operative, e.g.,
$gg\leftrightarrow ggg$
processes~\cite{Wong:1996,Xu:2004mz}. In this respect,
charm quarks are of particular importance, as their number is
presumably frozen after primordial production, and genuine 2$\to$3
processes are absent. Indeed, the recent analysis of
Ref.~\cite{vanHees:2004gq} has shown that ``$D$"-meson resonances in
the QGP can accelerate thermal relaxation times obtained from pQCD
by a factor of $\sim$3. A rather straightforward extension of our
approach to the heavy-light sector should therefore be pursued. The
formation of mesonic composites in the cooling QGP phase of a
heavy-ion collision could furthermore serve as a
``pre-hadronization" mechanism, and thus improve phenomenologically
successful quark-coalescence models at RHIC
\cite{Fries:2004hd,Greco:2003xt,Hwa:2002tu} (\eg, with respect to
the question of energy conservation). Significant future efforts
will be required to possibly develop such a scheme into a
quantitative phenomenology. Further progress will also reside on
increasing information from finite-$T$ lattice QCD to provide both
input and constraints to a many-body approach as presented here.
Clearly, a thorough understanding of the intricate properties of the
strongly interacting matter above $T_c$, and its implications for
ultrarelativistic heavy-ion experiments, is an exciting future task.

\vspace{0.8cm}

\noindent
{\bf Acknowledgement} \\
We are grateful to P. Petreczky for sending us the data files of the
unquenched lattice results for the free energies used in this work.
One of us (RR) has been supported in part by a U.S. National Science
Foundation CAREER award under grant PHY-0449489. The work of MM has
been supported in part by U.S. Department of Energy (D.O.E.) under
cooperative research agreement \#DE-FC02-94ER40818.


\begin{thebibliography}{50}
\expandafter\ifx\csname
natexlab\endcsname\relax\def\natexlab#1{#1}\fi
\expandafter\ifx\csname bibnamefont\endcsname\relax
  \def\bibnamefont#1{#1}\fi
\expandafter\ifx\csname bibfnamefont\endcsname\relax
  \def\bibfnamefont#1{#1}\fi
\expandafter\ifx\csname citenamefont\endcsname\relax
  \def\citenamefont#1{#1}\fi
\expandafter\ifx\csname url\endcsname\relax
  \def\url#1{\texttt{#1}}\fi
\expandafter\ifx\csname urlprefix\endcsname\relax\def\urlprefix{URL
}\fi \providecommand{\bibinfo}[2]{#2}
\providecommand{\eprint}[2][]{\url{#2}}

\bibitem[{\citenamefont{Baier et~al.}(2001)\citenamefont{Baier, Mueller,
  Schiff, and Son}}]{Baier:2000sb}
\bibinfo{author}{\bibfnamefont{R.}~\bibnamefont{Baier}},
  \bibinfo{author}{\bibfnamefont{A.~H.} \bibnamefont{Mueller}},
  \bibinfo{author}{\bibfnamefont{D.}~\bibnamefont{Schiff}}, \bibnamefont{and}
  \bibinfo{author}{\bibfnamefont{D.~T.} \bibnamefont{Son}},
  \bibinfo{journal}{Phys. Lett.} \textbf{\bibinfo{volume}{B502}},
  \bibinfo{pages}{51} (\bibinfo{year}{2001}).

\bibitem[{\citenamefont{Kolb and Heinz}(2003)}]{Kolb:2003dz}
\bibinfo{author}{\bibfnamefont{P.~F.} \bibnamefont{Kolb}} \bibnamefont{and}
  \bibinfo{author}{\bibfnamefont{U.}~\bibnamefont{Heinz}}
  (\bibinfo{year}{2003}), \eprint{nucl-th/0305084}.

\bibitem[{\citenamefont{Teaney et~al.}(2001)\citenamefont{Teaney, Lauret, and
  Shuryak}}]{Teaney:2000cw}
\bibinfo{author}{\bibfnamefont{D.}~\bibnamefont{Teaney}},
  \bibinfo{author}{\bibfnamefont{J.}~\bibnamefont{Lauret}}, \bibnamefont{and}
  \bibinfo{author}{\bibfnamefont{E.~V.} \bibnamefont{Shuryak}},
  \bibinfo{journal}{Phys. Rev. Lett.} \textbf{\bibinfo{volume}{86}},
  \bibinfo{pages}{4783} (\bibinfo{year}{2001}).

\bibitem[{\citenamefont{Hirano}(2004)}]{Hirano:2004er}
\bibinfo{author}{\bibfnamefont{T.}~\bibnamefont{Hirano}}, \bibinfo{journal}{J.
  Phys.} \textbf{\bibinfo{volume}{G30}}, \bibinfo{pages}{S845}
  (\bibinfo{year}{2004}).

\bibitem[{\citenamefont{Datta et~al.}(2003)\citenamefont{Datta, Karsch,
  Petreczky, and Wetzorke}}]{Datta:2002ck}
\bibinfo{author}{\bibfnamefont{S.}~\bibnamefont{Datta}},
  \bibinfo{author}{\bibfnamefont{F.}~\bibnamefont{Karsch}},
  \bibinfo{author}{\bibfnamefont{P.}~\bibnamefont{Petreczky}},
  \bibnamefont{and} \bibinfo{author}{\bibfnamefont{I.}~\bibnamefont{Wetzorke}},
  \bibinfo{journal}{Nucl. Phys. Proc. Suppl.} \textbf{\bibinfo{volume}{119}},
  \bibinfo{pages}{487} (\bibinfo{year}{2003}).

\bibitem[{\citenamefont{Asakawa and Hatsuda}(2004)}]{Asakawa:2003re}
\bibinfo{author}{\bibfnamefont{M.}~\bibnamefont{Asakawa}} \bibnamefont{and}
  \bibinfo{author}{\bibfnamefont{T.}~\bibnamefont{Hatsuda}},
  \bibinfo{journal}{Phys. Rev. Lett.} \textbf{\bibinfo{volume}{92}},
  \bibinfo{pages}{012001} (\bibinfo{year}{2004}), \eprint{hep-lat/0308034}.

\bibitem{Umeda:2002vr}
  T.~Umeda, K.~Nomura and H.~Matsufuru,
  Eur.\ Phys.\ J.\ C {\bf 39S1}, 9 (2005).

\bibitem[{\citenamefont{Karsch and Laermann}(2003)}]{Karsch:2003jg}
\bibinfo{author}{\bibfnamefont{F.}~\bibnamefont{Karsch}} \bibnamefont{and}
  \bibinfo{author}{\bibfnamefont{E.}~\bibnamefont{Laermann}}
  (\bibinfo{year}{2003}), \eprint{hep-lat/0305025}.

\bibitem[{\citenamefont{Asakawa and Hatsuda}(2003)}]{Asakawa:2003nw}
\bibinfo{author}{\bibfnamefont{M.}~\bibnamefont{Asakawa}} \bibnamefont{and}
  \bibinfo{author}{\bibfnamefont{T.}~\bibnamefont{Hatsuda}},
  \bibinfo{journal}{Nucl. Phys.} \textbf{\bibinfo{volume}{A721}},
  \bibinfo{pages}{869} (\bibinfo{year}{2003}).

\bibitem[{\citenamefont{van Hees and Rapp}(2004)}]{vanHees:2004gq}
\bibinfo{author}{\bibfnamefont{H.}~\bibnamefont{van Hees}} \bibnamefont{and}
  \bibinfo{author}{\bibfnamefont{R.}~\bibnamefont{Rapp}}
  \bibinfo{journal}{Phys. Rev. C} \textbf{\bibinfo{volume}{71}},
  \bibinfo{pages}{034907} (\bibinfo{year}{2005}).

\bibitem[{\citenamefont{Hatsuda and Kunihiro}(1984)}]{Hatsuda:1984jm}
\bibinfo{author}{\bibfnamefont{T.}~\bibnamefont{Hatsuda}} \bibnamefont{and}
  \bibinfo{author}{\bibfnamefont{T.}~\bibnamefont{Kunihiro}},
  \bibinfo{journal}{Phys. Lett.} \textbf{\bibinfo{volume}{B145}},
  \bibinfo{pages}{7} (\bibinfo{year}{1984}).

\bibitem[{\citenamefont{Hatsuda and Kunihiro}(1985)}]{Hatsuda:1985eb}
\bibinfo{author}{\bibfnamefont{T.}~\bibnamefont{Hatsuda}} \bibnamefont{and}
  \bibinfo{author}{\bibfnamefont{T.}~\bibnamefont{Kunihiro}},
  \bibinfo{journal}{Phys. Rev. Lett.} \textbf{\bibinfo{volume}{55}},
  \bibinfo{pages}{158} (\bibinfo{year}{1985}).

\bibitem[{\citenamefont{Schafer and Shuryak}(1995)}]{Schafer:1995df}
\bibinfo{author}{\bibfnamefont{T.}~\bibnamefont{Sch\"afer}} \bibnamefont{and}
  \bibinfo{author}{\bibfnamefont{E.~V.} \bibnamefont{Shuryak}},
  \bibinfo{journal}{Phys. Lett.} \textbf{\bibinfo{volume}{B356}},
  \bibinfo{pages}{147} (\bibinfo{year}{1995}).

\bibitem[{\citenamefont{Li et~al.}(2004)\citenamefont{Li, Li, Shakin, and
  Sun}}]{Li:2004ue}
\bibinfo{author}{\bibfnamefont{X.}~\bibnamefont{Li}},
  \bibinfo{author}{\bibfnamefont{H.}~\bibnamefont{Li}},
  \bibinfo{author}{\bibfnamefont{C.~M.} \bibnamefont{Shakin}},
  \bibnamefont{and} \bibinfo{author}{\bibfnamefont{Q.}~\bibnamefont{Sun}},
  \bibinfo{journal}{Phys. Rev. C} \textbf{\bibinfo{volume}{69}},
  \bibinfo{pages}{065201} (\bibinfo{year}{2004}).

\bibitem[{\citenamefont{Alberico et~al.}(2005)\citenamefont{Alberico, Beraudo,
  and Molinari}}]{Alberico:2004we}
\bibinfo{author}{\bibfnamefont{W.~M.} \bibnamefont{Alberico}},
  \bibinfo{author}{\bibfnamefont{A.}~\bibnamefont{Beraudo}}, \bibnamefont{and}
  \bibinfo{author}{\bibfnamefont{A.}~\bibnamefont{Molinari}},
  \bibinfo{journal}{Nucl. Phys.} \textbf{\bibinfo{volume}{A750}},
  \bibinfo{pages}{359} (\bibinfo{year}{2005}).

\bibitem[{\citenamefont{Digal et~al.}(2001)\citenamefont{Digal, Petreczky, and
  Satz}}]{Digal:2001ue}
\bibinfo{author}{\bibfnamefont{S.}~\bibnamefont{Digal}},
  \bibinfo{author}{\bibfnamefont{P.}~\bibnamefont{Petreczky}},
  \bibnamefont{and} \bibinfo{author}{\bibfnamefont{H.}~\bibnamefont{Satz}},
  \bibinfo{journal}{Phys. Rev. D} \textbf{\bibinfo{volume}{64}},
  \bibinfo{pages}{094015} (\bibinfo{year}{2001}).

\bibitem[{\citenamefont{Wong}(2004)}]{Wong:2004zr}
\bibinfo{author}{\bibfnamefont{C.-Y.} \bibnamefont{Wong}}
  (\bibinfo{year}{2004}), \eprint{hep-ph/0408020}.

\bibitem[{\citenamefont{Mocsy and Petreczky}(2004)}]{Mocsy:2004bv}
\bibinfo{author}{\bibfnamefont{A.}~\bibnamefont{Mocsy}} \bibnamefont{and}
  \bibinfo{author}{\bibfnamefont{P.}~\bibnamefont{Petreczky}}
  (\bibinfo{year}{2004}), \eprint{hep-ph/0411262}.

\bibitem[{\citenamefont{Shuryak and
  Zahed}(2004{\natexlab{a}})}]{Shuryak:2003ty}
\bibinfo{author}{\bibfnamefont{E.~V.} \bibnamefont{Shuryak}} \bibnamefont{and}
  \bibinfo{author}{\bibfnamefont{I.}~\bibnamefont{Zahed}},
  \bibinfo{journal}{Phys. Rev. C} \textbf{\bibinfo{volume}{70}},
  \bibinfo{pages}{021901} (\bibinfo{year}{2004}{\natexlab{a}}).

\bibitem[{\citenamefont{Brown et~al.}(2004)\citenamefont{Brown, Lee, Rho, and
  Shuryak}}]{Brown:2003km}
\bibinfo{author}{\bibfnamefont{G.~E.} \bibnamefont{Brown}},
  \bibinfo{author}{\bibfnamefont{C.-H.} \bibnamefont{Lee}},
  \bibinfo{author}{\bibfnamefont{M.}~\bibnamefont{Rho}}, \bibnamefont{and}
  \bibinfo{author}{\bibfnamefont{E.}~\bibnamefont{Shuryak}},
  \bibinfo{journal}{Nucl. Phys.} \textbf{\bibinfo{volume}{A740}},
  \bibinfo{pages}{171} (\bibinfo{year}{2004}).

\bibitem[{\citenamefont{Shuryak and
  Zahed}(2004{\natexlab{b}})}]{Shuryak:2004tx}
\bibinfo{author}{\bibfnamefont{E.~V.} \bibnamefont{Shuryak}} \bibnamefont{and}
  \bibinfo{author}{\bibfnamefont{I.}~\bibnamefont{Zahed}},
  \bibinfo{journal}{Phys. Rev. D} \textbf{\bibinfo{volume}{70}},
  \bibinfo{pages}{054507} (\bibinfo{year}{2004}{\natexlab{b}}).

\bibitem[{\citenamefont{Petreczky et~al.}(2002)\citenamefont{Petreczky, Karsch,
  Laermann, Stickan, and Wetzorke}}]{Petreczky:2001yp}
\bibinfo{author}{\bibfnamefont{P.}~\bibnamefont{Petreczky}},
  \bibinfo{author}{\bibfnamefont{F.}~\bibnamefont{Karsch}},
  \bibinfo{author}{\bibfnamefont{E.}~\bibnamefont{Laermann}},
  \bibinfo{author}{\bibfnamefont{S.}~\bibnamefont{Stickan}}, \bibnamefont{and}
  \bibinfo{author}{\bibfnamefont{I.}~\bibnamefont{Wetzorke}},
  \bibinfo{journal}{Nucl. Phys. Proc. Suppl.} \textbf{\bibinfo{volume}{106}},
  \bibinfo{pages}{513} (\bibinfo{year}{2002}).

\bibitem[{\citenamefont{Philipsen}(2002)}]{Philipsen:2002az}
\bibinfo{author}{\bibfnamefont{O.}~\bibnamefont{Philipsen}},
  \bibinfo{journal}{Phys. Lett.} \textbf{\bibinfo{volume}{B535}},
  \bibinfo{pages}{138} (\bibinfo{year}{2002}).

\bibitem[{\citenamefont{Kaczmarek et~al.}(2002)\citenamefont{Kaczmarek, Karsch,
  Petreczky, and Zantow}}]{Kaczmarek:2002mc}
\bibinfo{author}{\bibfnamefont{O.}~\bibnamefont{Kaczmarek}},
  \bibinfo{author}{\bibfnamefont{F.}~\bibnamefont{Karsch}},
  \bibinfo{author}{\bibfnamefont{P.}~\bibnamefont{Petreczky}},
  \bibnamefont{and} \bibinfo{author}{\bibfnamefont{F.}~\bibnamefont{Zantow}},
  \bibinfo{journal}{Phys. Lett.} \textbf{\bibinfo{volume}{B543}},
  \bibinfo{pages}{41} (\bibinfo{year}{2002}).

\bibitem[{\citenamefont{Kaczmarek
  et~al.}(2004{\natexlab{a}})\citenamefont{Kaczmarek, Karsch, Zantow, and
  Petreczky}}]{Kaczmarek:2004gv}
\bibinfo{author}{\bibfnamefont{O.}~\bibnamefont{Kaczmarek}},
  \bibinfo{author}{\bibfnamefont{F.}~\bibnamefont{Karsch}},
  \bibinfo{author}{\bibfnamefont{F.}~\bibnamefont{Zantow}}, \bibnamefont{and}
  \bibinfo{author}{\bibfnamefont{P.}~\bibnamefont{Petreczky}},
  \bibinfo{journal}{Phys. Rev. D} \textbf{\bibinfo{volume}{70}},
  \bibinfo{pages}{074505} (\bibinfo{year}{2004}{\natexlab{a}}).

\bibitem[{\citenamefont{Kaczmarek
  et~al.}(2004{\natexlab{b}})\citenamefont{Kaczmarek, Karsch, Petreczky, and
  Zantow}}]{Kaczmarek:2003dp}
\bibinfo{author}{\bibfnamefont{O.}~\bibnamefont{Kaczmarek}},
  \bibinfo{author}{\bibfnamefont{F.}~\bibnamefont{Karsch}},
  \bibinfo{author}{\bibfnamefont{P.}~\bibnamefont{Petreczky}},
  \bibnamefont{and} \bibinfo{author}{\bibfnamefont{F.}~\bibnamefont{Zantow}},
  \bibinfo{journal}{Nucl. Phys. Proc. Suppl.} \textbf{\bibinfo{volume}{129}},
  \bibinfo{pages}{560} (\bibinfo{year}{2004}{\natexlab{b}}).

\bibitem[{\citenamefont{Petreczky}(2004)}]{Petreczky:2004priv}
\bibinfo{author}{\bibfnamefont{P.}~\bibnamefont{Petreczky}},
  \bibinfo{journal}{private communcation}  (\bibinfo{year}{2004}).

\bibitem[{\citenamefont{Karsch et~al.}(1988)\citenamefont{Karsch, Mehr, and
  Satz}}]{Karsch:1987pv}
\bibinfo{author}{\bibfnamefont{F.}~\bibnamefont{Karsch}},
  \bibinfo{author}{\bibfnamefont{M.~T.} \bibnamefont{Mehr}}, \bibnamefont{and}
  \bibinfo{author}{\bibfnamefont{H.}~\bibnamefont{Satz}}, \bibinfo{journal}{Z.
  Phys.} \textbf{\bibinfo{volume}{C37}}, \bibinfo{pages}{617}
  (\bibinfo{year}{1988}).

\bibitem[{\citenamefont{Petreczky and Petrov}(2004)}]{Petreczky:2004pz}
\bibinfo{author}{\bibfnamefont{P.}~\bibnamefont{Petreczky}} \bibnamefont{and}
  \bibinfo{author}{\bibfnamefont{K.}~\bibnamefont{Petrov}},
  \bibinfo{journal}{Phys. Rev. D} \textbf{\bibinfo{volume}{70}},
  \bibinfo{pages}{054503} (\bibinfo{year}{2004}).

\bibitem[{\citenamefont{Digal}(2004)}]{Digal:2004}
\bibinfo{author}{\bibfnamefont{S.}~\bibnamefont{Digal}}, 
\bibinfo{journal}{Proceedings of the Int. Conference on ``Hard and
 Electromagnetic Probes of High-Energy Heavy-Ion Collisions" (Ericeira, 
 Portugal, Nov. 4-10, 2004)}  (\bibinfo{year}{2004}),
  \eprint{hep-ph/0505193}.

\bibitem[{\citenamefont{Jahn and Philipsen}(2004)}]{Jahn:2004qr}
\bibinfo{author}{\bibfnamefont{O.}~\bibnamefont{Jahn}} \bibnamefont{and}
  \bibinfo{author}{\bibfnamefont{O.}~\bibnamefont{Philipsen}},
  \bibinfo{journal}{Phys. Rev. D} \textbf{\bibinfo{volume}{70}},
  \bibinfo{pages}{074504} (\bibinfo{year}{2004}).

\bibitem[{\citenamefont{Brown}(1952)}]{Brown:1952ph}
\bibinfo{author}{\bibfnamefont{G.~E.} \bibnamefont{Brown}},
  \bibinfo{journal}{Philos. Mag.} \textbf{\bibinfo{volume}{43}},
  \bibinfo{pages}{467} (\bibinfo{year}{1952}).

\bibitem[{\citenamefont{Levai and Heinz}(1998)}]{Levai:1997yx}
\bibinfo{author}{\bibfnamefont{P.}~\bibnamefont{Levai}} \bibnamefont{and}
  \bibinfo{author}{\bibfnamefont{U.~W.} \bibnamefont{Heinz}},
  \bibinfo{journal}{Phys. Rev. C} \textbf{\bibinfo{volume}{57}},
  \bibinfo{pages}{1879} (\bibinfo{year}{1998}).

\bibitem[{\citenamefont{Schneider and Weise}(2001)}]{Schneider:2001nf}
\bibinfo{author}{\bibfnamefont{R.A.} \bibnamefont{Schneider}}
  \bibnamefont{and} \bibinfo{author}{\bibfnamefont{W.}~\bibnamefont{Weise}},
  \bibinfo{journal}{Phys. Rev. C} \textbf{\bibinfo{volume}{64}},
  \bibinfo{pages}{055201} (\bibinfo{year}{2001}).

\bibitem[{\citenamefont{Peshier et~al.}(2002)\citenamefont{Peshier, Kampfer,
  and Soff}}]{Peshier:2002ww}
\bibinfo{author}{\bibfnamefont{A.}~\bibnamefont{Peshier}},
  \bibinfo{author}{\bibfnamefont{B.}~\bibnamefont{Kampfer}}, \bibnamefont{and}
  \bibinfo{author}{\bibfnamefont{G.}~\bibnamefont{Soff}},
  \bibinfo{journal}{Phys. Rev. D} \textbf{\bibinfo{volume}{66}},
  \bibinfo{pages}{094003} (\bibinfo{year}{2002}).

\bibitem[{\citenamefont{Blaizot et~al.}(2003)\citenamefont{Blaizot, Iancu, and
  Rebhan}}]{Blaizot:2003tw}
\bibinfo{author}{\bibfnamefont{J.-P.} \bibnamefont{Blaizot}},
  \bibinfo{author}{\bibfnamefont{E.}~\bibnamefont{Iancu}}, \bibnamefont{and}
  \bibinfo{author}{\bibfnamefont{A.}~\bibnamefont{Rebhan}}
  (\bibinfo{year}{2003}), \eprint{hep-ph/0303185}.

\bibitem[{\citenamefont{Kraemmer and Rebhan}(2004)}]{Kraemmer:2003gd}
\bibinfo{author}{\bibfnamefont{U.}~\bibnamefont{Kraemmer}} \bibnamefont{and}
  \bibinfo{author}{\bibfnamefont{A.}~\bibnamefont{Rebhan}},
  \bibinfo{journal}{Rept. Prog. Phys.} \textbf{\bibinfo{volume}{67}},
  \bibinfo{pages}{351} (\bibinfo{year}{2004}).

\bibitem[{\citenamefont{Blankenbecler and Sugar}(1966)}]{Blankenbecler:1965gx}
\bibinfo{author}{\bibfnamefont{R.}~\bibnamefont{Blankenbecler}}
  \bibnamefont{and} \bibinfo{author}{\bibfnamefont{R.}~\bibnamefont{Sugar}},
  \bibinfo{journal}{Phys. Rev.} \textbf{\bibinfo{volume}{142}},
  \bibinfo{pages}{1051} (\bibinfo{year}{1966}).

\bibitem[{\citenamefont{Thompson}(1970)}]{Thompson:1970wt}
\bibinfo{author}{\bibfnamefont{R.~H.} \bibnamefont{Thompson}},
  \bibinfo{journal}{Phys. Rev. D} \textbf{\bibinfo{volume}{1}},
  \bibinfo{pages}{110} (\bibinfo{year}{1970}).

\bibitem[{\citenamefont{Asakawa et~al.}(2003)\citenamefont{Asakawa, Hatsuda,
  and Nakahara}}]{Asakawa:2002xj}
\bibinfo{author}{\bibfnamefont{M.}~\bibnamefont{Asakawa}},
  \bibinfo{author}{\bibfnamefont{T.}~\bibnamefont{Hatsuda}}, \bibnamefont{and}
  \bibinfo{author}{\bibfnamefont{Y.}~\bibnamefont{Nakahara}},
  \bibinfo{journal}{Nucl. Phys.} \textbf{\bibinfo{volume}{A715}},
  \bibinfo{pages}{863} (\bibinfo{year}{2003}).

\bibitem[{\citenamefont{Haftel and Tabakin}(1970)}]{Haftel:1970}
\bibinfo{author}{\bibfnamefont{M.~I.} \bibnamefont{Haftel}} \bibnamefont{and}
  \bibinfo{author}{\bibfnamefont{F.}~\bibnamefont{Tabakin}},
  \bibinfo{journal}{Nucl. Phys.} \textbf{\bibinfo{volume}{A158}},
  \bibinfo{pages}{1} (\bibinfo{year}{1970}).

\bibitem{Grandchamp:2003uw}
  L.~Grandchamp, R.~Rapp and G.E.~Brown,
  Phys.\ Rev.\ Lett.\  {\bf 92}, 212301 (2004).

\bibitem[{\citenamefont{Peshier and Cassing}(2005)}]{Peshier:2005pp}
\bibinfo{author}{\bibfnamefont{A.}~\bibnamefont{Peshier}} \bibnamefont{and}
  \bibinfo{author}{\bibfnamefont{W.}~\bibnamefont{Cassing}}
  (\bibinfo{year}{2005}), \eprint{hep-ph/0502138}.

\bibitem[{\citenamefont{Thoma}(2005)}]{Thoma:2005uv}
\bibinfo{author}{\bibfnamefont{M.~H.} \bibnamefont{Thoma}},
  \bibinfo{journal}{J. Phys.} \textbf{\bibinfo{volume}{G31}},
  \bibinfo{pages}{L7} (\bibinfo{year}{2005}).

\bibitem[{\citenamefont{Rapp and Wambach}(1995)}]{Rapp:1995fv}
\bibinfo{author}{\bibfnamefont{R.}~\bibnamefont{Rapp}} \bibnamefont{and}
  \bibinfo{author}{\bibfnamefont{J.}~\bibnamefont{Wambach}},
  \bibinfo{journal}{Phys. Lett.} \textbf{\bibinfo{volume}{B351}},
  \bibinfo{pages}{50} (\bibinfo{year}{1995}).

\bibitem[{\citenamefont{Wong}(1996{\natexlab{a}})}]{Wong:1996}
\bibinfo{author}{\bibfnamefont{S.~M.~H.} \bibnamefont{Wong}},
  \bibinfo{journal}{Nucl. Phys.} \textbf{\bibinfo{volume}{A607}},
  \bibinfo{pages}{442} (\bibinfo{year}{1996}{\natexlab{a}});
 \bibinfo{journal}{Phys. Rev. C} \textbf{\bibinfo{volume}{54}},
  \bibinfo{pages}{2588} (\bibinfo{year}{1996}{\natexlab{b}}).

\bibitem[{\citenamefont{Xu and Greiner}(2004)}]{Xu:2004mz}
\bibinfo{author}{\bibfnamefont{Z.}~\bibnamefont{Xu}} \bibnamefont{and}
  \bibinfo{author}{\bibfnamefont{C.}~\bibnamefont{Greiner}}
  (\bibinfo{year}{2004}), \eprint{hep-ph/0406278}.

\bibitem[{\citenamefont{Fries et~al.}(2005)\citenamefont{Fries, Bass, and
  Muller}}]{Fries:2004hd}
\bibinfo{author}{\bibfnamefont{R.~J.} \bibnamefont{Fries}},
  \bibinfo{author}{\bibfnamefont{S.~A.} \bibnamefont{Bass}}, \bibnamefont{and}
  \bibinfo{author}{\bibfnamefont{B.}~\bibnamefont{Muller}},
  \bibinfo{journal}{Phys. Rev. Lett.} \textbf{\bibinfo{volume}{94}},
  \bibinfo{pages}{122301} (\bibinfo{year}{2005}).

\bibitem[{\citenamefont{Greco et~al.}(2003)\citenamefont{Greco, Ko, and
  Levai}}]{Greco:2003xt}
\bibinfo{author}{\bibfnamefont{V.}~\bibnamefont{Greco}},
  \bibinfo{author}{\bibfnamefont{C.~M.} \bibnamefont{Ko}}, \bibnamefont{and}
  \bibinfo{author}{\bibfnamefont{P.}~\bibnamefont{Levai}},
  \bibinfo{journal}{Phys. Rev. Lett.} \textbf{\bibinfo{volume}{90}},
  \bibinfo{pages}{202302} (\bibinfo{year}{2003}).

\bibitem[{\citenamefont{Hwa and Yang}(2003)}]{Hwa:2002tu}
\bibinfo{author}{\bibfnamefont{R.~C.} \bibnamefont{Hwa}} \bibnamefont{and}
  \bibinfo{author}{\bibfnamefont{C.~B.} \bibnamefont{Yang}},
  \bibinfo{journal}{Phys. Rev. C} \textbf{\bibinfo{volume}{67}},
  \bibinfo{pages}{034902} (\bibinfo{year}{2003}).

\end{thebibliography}

\end{document}